\documentclass[aps,prb,twocolumn,superscriptaddress,10pt]{revtex4}
\usepackage{amsmath}
\usepackage{amssymb}
\usepackage{graphicx}
\usepackage{subcaption}
\usepackage{epsfig}
\usepackage{soul} 
\usepackage{xcolor}
\usepackage{caption}
\usepackage{blindtext}
\usepackage{natbib}
\usepackage{color}
\usepackage{mathtools}

\def\be{\begin{equation}} \def\ee{\end{equation}}
\def\bea{\begin{eqnarray}} \def\eea{\end{eqnarray}}

\def\nn{\nonumber}

\definecolor{mycolor}{RGB}{255,255,0}
\sethlcolor{mycolor}

\begin{document}
\title{Tri-component-pairing chiral superconductivity on the honeycomb lattice with mixed $s$- and $d$-wave symmetries}

\author{Yu-Hang Li}
\affiliation{School of Physics, Nankai University, Tianjin 300071, China}

\author{Jiarui Jiao}
\affiliation{School of Physics, Nankai University, Tianjin 300071, China}

\author{Xiao-Xiao Zhang}
\affiliation{Wuhan National High Magnetic Field Center and School of Physics,
Huazhong University of Science and Technology, Wuhan 430074, China}

\author{Congjun Wu}
\email{wucongjun@westlake.edu.cn}
\affiliation{New Cornerstone Science Laboratory, Department of Physics, School of Science, Westlake University, Hangzhou 310024, Zhejiang, China}
\affiliation{Institute for Theoretical Sciences, Westlake University, Hangzhou 310024, Zhejiang, China}
\affiliation{Key Laboratory for Quantum Materials of Zhejiang Province, School of Science, Westlake University, Hangzhou 310024, Zhejiang, China}
\affiliation{Institute of Natural Sciences, Westlake Institute for Advanced Study, Hangzhou 310024, Zhejiang, China}

\author{Wang Yang}
\email{wyang@nankai.edu.cn}
\affiliation{School of Physics, Nankai University, Tianjin 300071, China}

\begin{abstract}
In this work, we investigate chiral topological superconductors on a two-dimensional honeycomb lattice with coexisting $d_{x^2-y^2}$, $d_{xy}$, and $s$-wave pairing symmetries. 
Using a Ginzburg-Landau free energy analysis, the pairing gap function is shown to exhibit a tri-component form $s + d_{x^2-y^2} e^{i\phi_1} + d_{xy} e^{i\phi_2}$, where $\phi_1$ and $\phi_2$ are phase differences between the $d$- and $s$-wave pairing components,
which spontaneously breaks both time reversal and $C_6$ rotational symmetries. 
Chern numbers of the energy bands are calculated to be nonzero, demonstrating the topologically nontrivial nature of the system.
The anomalous AC Hall conductivity is  computed, which is not invariant under $C_6$ rotations, reflecting the anisotropic nature of the  pairing gap function.
Fractional magnetic vortices are also discussed, arising from the multi-component nature of the pairing gap function. 
\end{abstract}

\maketitle
 

\section{introduction}

Chiral superconductors \cite{Kallin2016} have attracted intense research interests because of their potentials in realizing topological quantum computations \cite{Sarma2015,Freedman2003,Nayak2008,Lahtinen2017,Qi2010,Sato2017,zhang2018,frolov2020,Tanaka2024,Lutchyn2018,Sun2017,Flensberg2021,Mercado2022,Beenakker2013,Sato2016,Franz2010,Rahmani2019}. 
Typical chiral pairing states include chiral $p$-wave \cite{Furusaki2001,Yokoyama2008,Kallin2012,Nelson2004,Luke1998,Maeno2011,Mackenzie2017,Fittipaldi2021,Tummuru2021}, $d$-wave \cite{Fischer2014,Liu2017,Awoga2017,Kittaka2016,Can2021} and $f$-wave pairings \cite{Wang2016,Schemm2014,Sauls1994,Avers2020,Strand2010}, corresponding to orbital angular momentum of  a Cooper pair equal to one, two, and three, respectively.
Possible chiral $d$-wave superconductors include certain copper-oxide high-temperature superconductors \cite{Ito1991,Pickett1989,Tranquada2004,Keimer2015,Klemm1990} and honeycomb correlated systems \cite{Brydon2019,Wu2013,Black-Schaffer2014}. 
Based on the doped Hubbard model on the honeycomb lattice, it is theoretically proposed that superconductivity arising from correlated electrons may take the form of a chiral $d \pm id$ singlet pairing or a $p \pm ip$ triplet pairing, depending on the doping level and interaction strength \cite{Ying2020,Roy2010,Lin2018,Su2009}.
At doping levels near the van Hove singularity (VHS), the $d \pm id$ singlet pairing dominates at weak coupling,
while the $p \pm ip$ triplet pairing becomes more prominent when the interaction strength increases \cite{Xu2016}.
Recently, there has been evidence that the surface of YPtBi material may host chiral $d+id$ superconducting pairing \cite{Persky2025,Schwemmer2022}.

Exotic chiral superconductivities can also emerge from multi-layer systems. 
For example, recent studies have revealed the possible  presence of chiral $d$-wave superconducting pairing in twisted bilayer graphene \cite{Yankowitz2019,Tarnopolsky2019,Cao2020,Lisi2021,Kerelsky2019,Moon2013,Samajdar2020,Peltonen2018,Cao2021,WangYX2021}. 
Interestingly, it has been proposed in Ref. \onlinecite{Can2021} that the method of twisting bilayer materials provides a strategy to stack two layers of Bi2212 thin films together and rotate them at a certain angle.
When the twist angle increases from 0 to $\mathrm{45^\circ}$ in the twisted cuprate system, the pairing symmetry transitions from $s \pm id$ to  $d_{x^2-y^2} \pm id_{xy}$ \cite{Can2021,can2021,Yu2019}.
In addition, the study of a superconducting heterojunction with one side characterized by the $p_x \pm ip_y$ gap function and the other side the conventional $s$-wave one found the pairing pattern to be $s + i\eta_1[e^{i\eta_2\phi/2}p_x+\eta_3e^{-i\eta_2\phi/2}p_y]$ with $\eta_j=\pm1 (j =1,2,3)$, where $\phi$ is the phase difference between the $p_x$- and $p_y$-wave pairing components \cite{Xu2023}.

One significant feature of chiral superconductivity is the spontaneous breaking of time reversal symmetry,
as signified by the non-collinear phase difference between different components of pairing order parameters. 
Time-reversal symmetry breaking can be detected through various methods, such as muon spin relaxation experiments \cite{Luke1993,Uemura1985},
Josephson interference measurements \cite{Jaklevic1964,Giazotto2012,Gerace2009,Fulton1972,Li2019,Song2016},
the magneto-optical Faraday effect \cite{Ferreira2011},
and Kerr rotation experiments \cite{Ingersoll1912,Cebollada1994,Shimano2013,Mertins2004,Subkhangulov2016}.
In Kerr rotation experiments, the system breaking time-reversal symmetry typically exhibits a nonzero Kerr rotation angle of light, meaning that the polarization direction of reflected light rotates.
This effect can be measured using ultra-high sensitivity zero-field Kerr effect measurements.
Since the Kerr angle is related to the AC  Hall conductivity $\sigma_H$ under zero external magnetic field,
a non-vanishing AC Hall conductivity is an evidence for the existence of time reversal symmetry breaking in the system \cite{Tse2011,Marui2023}.

In this paper, we investigate chiral superconductors on a honeycomb lattice in the case where nearest-neighbor pairing dominates,
with coexisting $d_{x^2-y^2}$, $d_{xy}$, and $s$-wave pairing symmetries. 
The coexistence of these three pairing symmetries can be either intrinsic or extrinsic,
where ``intrinsic" refers to simultaneous  instabilities in the three superconducting channels in the material,
and ``extrinsic" refers to the situation where the coexistence is induced via proximity effect by superimposing a conventional $s$-wave superconductor on top of a chiral $d \pm id$ one.
The experimentally observed nematic superconductivity in twisted bilayer graphene \cite{Cao2021} has been proposed to be possibly arising from the intrinsic coexistence of these pairing components as discussed in Ref. \onlinecite{WangYX2021};
while the extrinsic scenario is applicable to any $d+id$ pairing material in proximity  with an $s$-wave one.
As to be discussed shortly, the $C_6$ rotational symmetry is spontaneously broken for the tri-component pairing,
whereas the $d \pm id$ pairing preserves this symmetry.
Hence, the resulting anisotropic Hall response signals can serve as signatures for nematic superconductivity in ``intrinsic" case,
and chiral $d \pm id$ pairing in the ``extrinsic"  case in the materials.

\begin{figure}[h]
\centering
\includegraphics[width=0.22\textwidth]{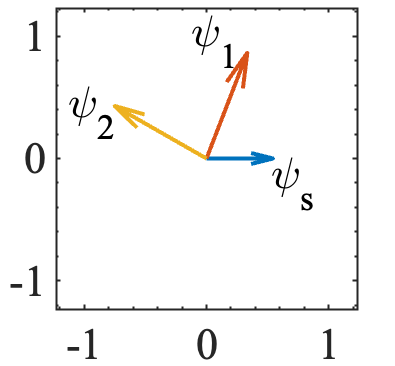}
 \captionsetup{justification=raggedright}
\caption{One of the twelve  degenerate configurations of the tri-component pairing gap function on honeycomb lattice,
in which $\psi_s$, $\psi_1$, and $\psi_2$ represent the $s$-, $d_{x^2-y^2}$-, and $d_{xy}$-pairing components, respectively. 
The phase of $s$-wave order parameter is fixed to zero, i.e., $|\psi_s| + |\psi_1| e^{i\phi_1} + |\psi_2| e^{i\phi_2}$.
The choices of the values of parameters in the free energy are included in the text. 
It is worth emphasizing that the two components $\psi_1$ and $\psi_2$ in the figure have a phase difference of $0.452\pi$,  not perpendicular with each other. }
\label{fig:C6vdds}
\end{figure}

From a free energy perspective, there are complex frustrations and intertwinings among the $d_{x^2-y^2}$-, $d_{xy}$-, and $s$-wave pairing components.  
On the one hand, the quadratic Josephson couplings favor a relative $\pm \pi/2$ phase difference between any two of the three pairing symmetries;
and on the other hand, the existence of an exotic quartic term in the free energy which is first order in $\psi_s$ and cubic in $d$-wave components
favors a phase difference of $n\pi$ (or $m\pi/2$) between the $d_{x^2-y^2}$- (or $d_{xy}$-) wave and $s$-wave pairings, where both $m$ and $n$ are integers.
However, the phase differences among the three pairing components cannot simultaneously satisfy all these conditions.
Based on a Ginzburg-Landau (GL) free energy analysis, we find that the pairing gap function is of the form $s+d_{x^2-y^2}e^{i\phi_1}+d_{xy}e^{i\phi_2}$, where $\phi_1$ (and $\phi_2$) represents the phase difference between the $d_{x^2-y^2}$- (and $d_{xy}$-) wave and the $s$-wave pairing order parameters as shown in Fig. \ref{fig:C6vdds}. 

The obtained pattern of tri-component pairing $s+d_{x^2-y^2}e^{i\phi_1}+d_{xy}e^{i\phi_2}$  not only spontaneously breaks the time-reversal symmetry,
but also breaks the spatial $C_{6v}$ symmetry of the honeycomb lattice down to $C_2$.
The breaking of time reversal symmetry manifests itself in the non-vanishing Hall conductivity,
whereas the absence of $C_6$ rotational symmetry in the pairing gap function can be detected through the spatial anisotropy in Hall conductivity and Kerr effect.
Furthermore, we have confirmed that the pairing is topologically nontrivial by 
showing the non-vanishing of Chern number and the emergence of Majorana edge mode on the boundaries. 

Because of the multi-component structure of the pairing gap function, 
the system can host exotic topological excitations,
not possible in superconductors with a single pairing component. 
In particular, we show in detail that the tri-component pairing superconductivity can host magnetic vortices carrying arbitrary fractions of the magnetic flux quantum \cite{Babaev2002}.
Other exotic topological excitations such as chiral skyrmions  can also exist in the tri-component pairing system \cite{Garaud2013}.  
Three-component superconductors have been studied to some extent, revealing
spontaneous time-reversal symmetry breaking \cite{Bojesen2013},
and novel topological solitons \cite{Garaud2011}.

The rest of the paper is organized as follows.
In Sec. \ref{sec:GL}, we begin with a GL free energy analysis, from which the form of the pairing gap function and the symmetry breaking pattern  are derived.
In Sec. \ref{sec:topo}, by using a microscopic Bogoliubov-de Gennes (BdG) Hamiltonian of a pairing gap function $s+d_{x^2-y^2}e^{i\phi_1}+d_{xy}e^{i\phi_2}$ on the honeycomb lattice,   we show the opening of the topological mass gap and the non-zero Chern number. 
The anisotropic anomalous AC Hall conductivity is studied  in Sec. \ref{sec:Hall}.
In Sec. \ref{sec:fractional_vortex}, fractional magnetic vortices are discussed. 
Conclusions are presented in Sec. \ref{sec:conclusion}.

\section{Ginzburg-Landau free energy analysis}
\label{sec:GL}

\subsection{Uniform Ginzburg–Landau free energy}
\label{subsec:free_energy_C6v}

\begin{figure}[h]
\centering
\includegraphics[width=0.245\textwidth]{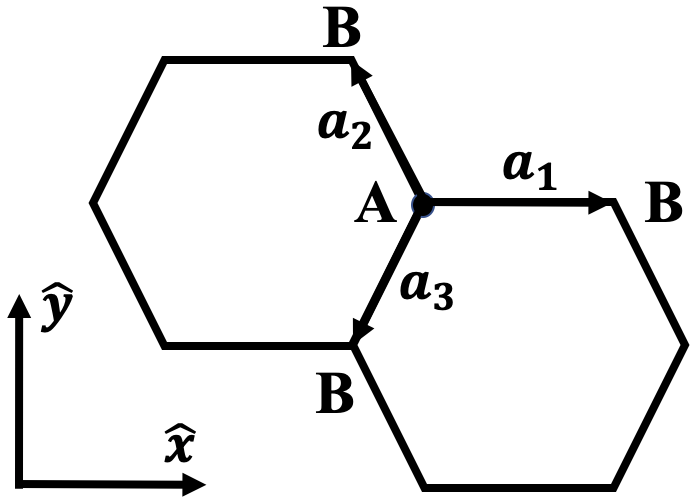}
 \captionsetup{justification=raggedright}
\caption{Schematic plot of a two-dimensional honeycomb lattice,
where $A$ and $B$ denote sites in the two inequivalent sublattices.
The three nearest-neighbour vectors for the sublattice site $A$ are shown as the black arrows as
$\boldsymbol{a_1}=(a,0)$, $\boldsymbol{a_2}=(-a/2,\sqrt{3}a/2)$, and $\boldsymbol{a_3}=(-a/2,-\sqrt{3}a/2)$,
in which the lattice constant of the honeycomb lattice is $a$.
The $x$-direction is taken as the direction pointing from sublattice site $A$ to $B$, and the $y$-direction is in the perpendicular direction.
}
\label{fig:nearestn}
\end{figure}

We consider a superconducting system on the honeycomb lattice as shown in Fig. \ref{fig:nearestn}. 
The superconducting pairing gap function will be shown to exhibit a tri-component form with competing $d_{x^2-y^2}$-, $d_{xy}$-, and $s$-wave pairing symmetries, based on a combination of symmetry and GL free energy analysis. 

\begin{figure}[h]
\centering
\includegraphics[width=0.3\textwidth]{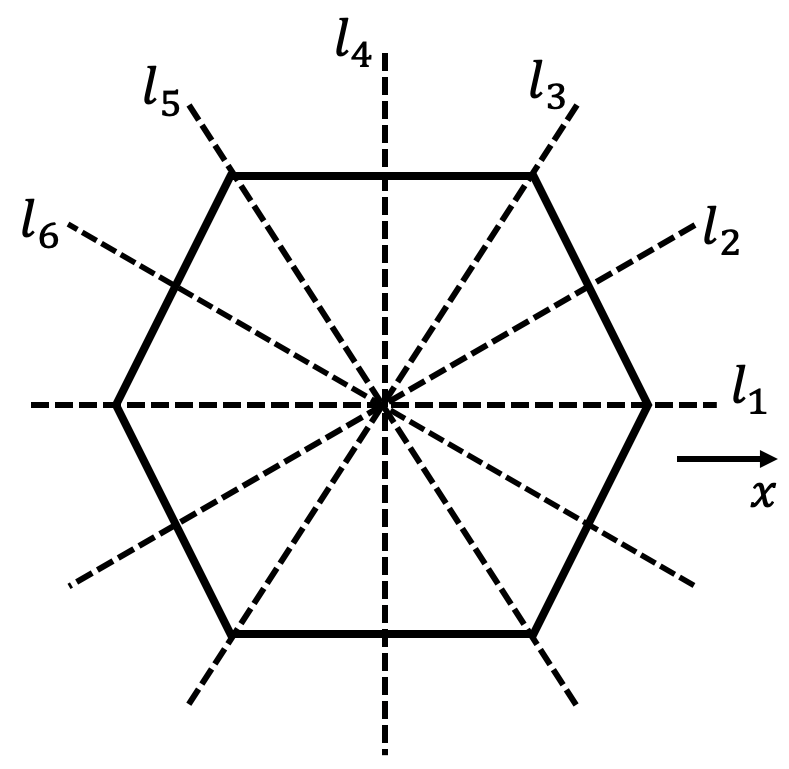}
 \captionsetup{justification=raggedright}
\caption{Schematic plot of the twelve symmetry elements of the $C_{6v}$ group consisting of six rotation  and six reflection operations.
The rotation operations are represented as $C_6$, $C_3$, $C_2$, $C_3^2$ and $C_6^5$ in the  text, corresponding to rotations around $z$-axis by angles $\pi/3$, $2\pi/3$, $\pi$, $4\pi/3$, and $5\pi/3$. 
The six reflection planes of the reflection operations are determined by the planes spanned by $z$-axis and the dashed lines $l_1$, $l_2$, $l_3$, $l_4$, $l_5$, $l_6$. }
\label{fig:C6vgroup}
\end{figure}

The point group symmetry of a monolayer of honeycomb lattice is $C_{6v}$, which contains six rotations and six reflections, as shown in Fig. \ref{fig:C6vgroup}.
The most general GL free energy respecting the $U(1)$ gauge, the time reversal, and the $C_{6v}$ point group symmetries up to the quartic order is given by:
\begin{eqnarray}
F=F_{s}^{(0)} + F_{d}^{(0)}+F^{(4)},
\label{eq:F}
\end{eqnarray}
in which
\begin{eqnarray}
F_{s}^{(0)}&=&\alpha_s |\psi_s|^2 + \beta_s |\psi_s|^4,\notag\\
F_{d}^{(0)}&=&\alpha_d \left( |\psi_1|^2+|\psi_2|^2 \right) + \beta_d \left( |\psi_1|^2+|\psi_2|^2 \right)^2,
\label{eq:F0}
\end{eqnarray}
and
\begin{align}
F^{(4)}\:&=\: \gamma |\psi_s|^2 \left( |\psi_1|^2+|\psi_2|^2 \right) +g_{dd} \left( \psi^*_1 \psi_2 - \psi_1 \psi^*_2 \right)^2 \notag\\
\:&+\:g_{sd} \left[ \psi_s^{*2} \left( \psi_1^2 + \psi_2^2 \right) + \psi_s^2 \left( \psi_1^{*2} + \psi_2^{*2} \right)\right] \notag\\
&\begin{aligned}
\:+\:\:g'_{sd}&\left[ \left( \psi_s^* \psi_1 + \psi_s \psi_1^* \right) \left(|\psi_1|^2-|\psi_2|^2 \right) \right. \\
&\left. -\left( \psi_s^* \psi_2 + \psi_s \psi_2^* \right) \left( \psi_1^*\psi_2+\psi_1\psi_2^* \right) \right],
\end{aligned}
\label{eq:F4}
\end{align}
where $\psi_s$, $\psi_1$ and $\psi_2$ represent the complex order parameters for the $s$-wave, $d_{x^2-y^2}$-wave and $d_{xy}$-wave, respectively;
$\alpha_s<0$, $\alpha_d<0$, $\beta_s>0$, $\beta_d>0$ in the superconducting phase when three pairing symmetries coexist;
$\gamma$ represents the phase-independent coupling term between the $s$-wave and $d$-wave pairing components;
$g_{dd}>0$ is the coefficient of the term which contains the quadratic Josephson coupling $\psi_1^2 \psi_2^{*2}+h.c.$ between $d_{x^2-y^2}$- and $d_{xy}$-wave components;
$g_{sd}>0$ is the coefficients of the quadratic Josephson coupling $\psi_s^{*2}(\psi_1^2+\psi_2^2)+h.c.$ between $s$- and $d$-wave components;
$g_{sd}^\prime$ represents the quartic coupling term which is first order in $\psi_s$ and cubic in $d$-wave components. 
We note that both $g_{sd}>0$ and $g_{dd}>0$ are taken to be positive so that relative $\pm \pi/2$ phase differences are energetically favored  between any two of the three pairing components $\psi_s,\psi_1,\psi_2$.
In what follows, by setting the phase of $\psi_s$ to zero, we write 
\begin{equation}
\psi_s = |\psi_s|, ~~
\psi_1 = |\psi_1| e^{i\phi_1}, ~~
\psi_2 = |\psi_2| e^{i\phi_2},
\label{eq:orderparameters}
\end{equation}
in which $|\psi_s|$, $|\psi_1|$ and $|\psi_2|$ are magnitudes of the $s$-wave, $d_{x^2-y^2}$-wave and $d_{xy}$-wave order parameters, 
and $\phi_1$ and $\phi_2$ are the phase differences of $\psi_1$ and $\psi_2$ relative to $\psi_s$.
Notice that it is the term with coefficient $g'_{sd}$ which breaks the $U(1)$ rotational  symmetry down to $C_{6v}$. 
A more detailed derivation of Eq. \eqref{eq:F} based on symmetry anaylsis is provided in Appendix \ref{app:glfree}.
The origin of the tri-component form of the pairing gap function can be most evidently seen 
by retaining only the phase-sensitive terms in Eq. \eqref{eq:F}.
Plugging the expressions of $\psi_s$, $\psi_1$ and $\psi_2$ in Eq. (\ref{eq:orderparameters}) into Eq. (\ref{eq:F}), we obtain
\begin{eqnarray}
F&=& f_1(|\psi_s|,|\psi_1|) \cos{2\phi_1} \notag\\
&+& f_2(|\psi_s|,|\psi_2|) \cos{2\phi_2} \notag\\
&+& f_0(|\psi_1|,|\psi_2|) \cos{(2\phi_2–2\phi_1)} \notag\\
&+& f'_1(|\psi_s|,|\psi_1|,|\psi_2|) \cos{\phi_1} \notag\\
&+& f'_2(|\psi_s|,|\psi_1|,|\psi_2|) \cos{(2\phi_2-\phi_1)},
\label{eq:Fphi}
\end{eqnarray}
where
\begin{align}
& f_1(|\psi_s|,|\psi_1|) = 2 g_{sd} |\psi_s|^2 |\psi_1|^2, \notag\\
& f_2(|\psi_s|,|\psi_2|) = 2 g_{sd} |\psi_s|^2 |\psi_2|^2, \notag\\
& f_0(|\psi_1|,|\psi_2|) = 2 g_{dd} |\psi_1|^2 |\psi_2|^2, \notag\\
& f'_1(|\psi_s|,|\psi_1|,|\psi_2|) = 2g'_{sd} |\psi_s| |\psi_1| \left( |\psi_1|^2-2|\psi_2|^2 \right), \notag\\
& f'_2(|\psi_s|,|\psi_1|,|\psi_2|) = -2g'_{sd} |\psi_s| |\psi_1| |\psi_2|^2.
\label{eq:f1f2f0}
\end{align}
Since $f_1$, $f_2$ and $f_0$ are all positive, $\phi_1$, $\phi_2$ and $\phi_2-\phi_1$ all tend to take values of $\pm \pi/2$,
meaning that at least one of $\phi_1$, $\phi_2$, or $\phi_2 - \phi_1$ will deviate from $\pm \pi/2$.

\begin{figure*}[htb]
    \centering
    \begin{subfigure}[b]{1\textwidth}
        \centering
        \includegraphics[width=\textwidth]{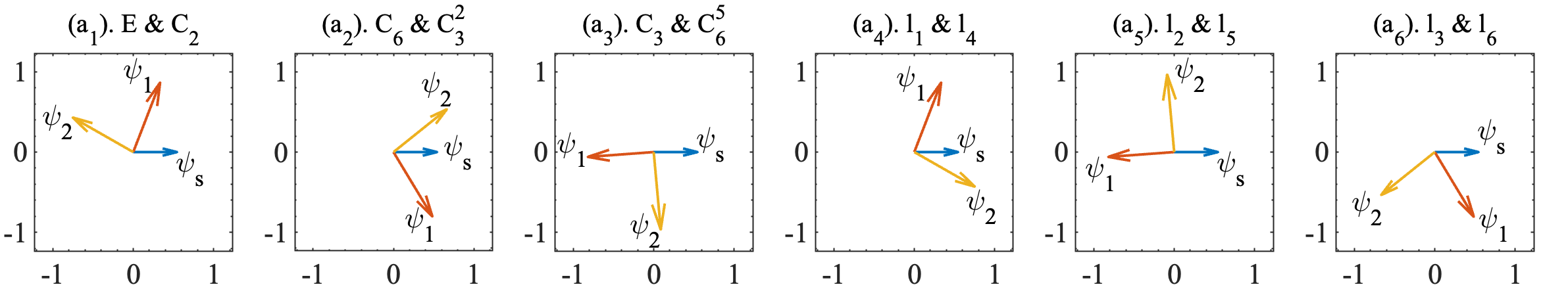}
        \caption{}
        \label{fig:subT0}
    \end{subfigure}
    \hfill
    \begin{subfigure}[b]{1\textwidth}
        \centering
        \includegraphics[width=\textwidth]{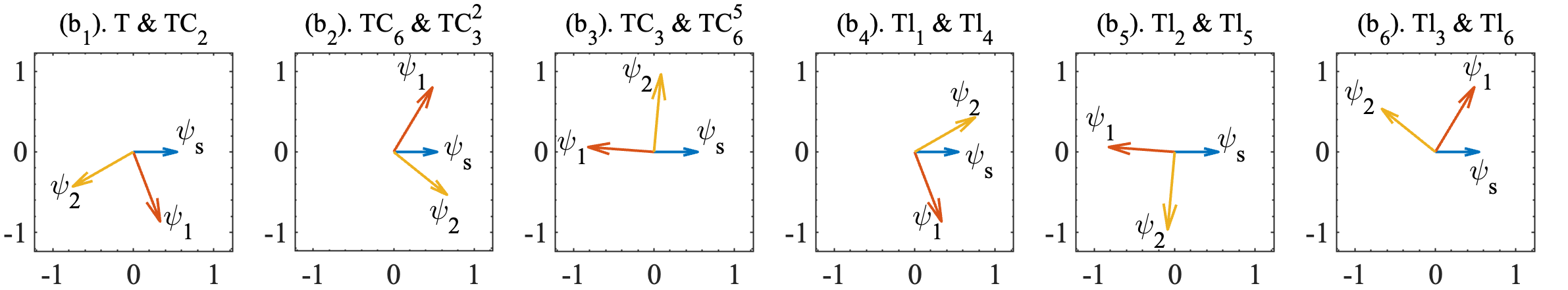}
        \caption{}
        \label{fig:subT1}
    \end{subfigure}
    \captionsetup{justification=raggedright}
    \caption{Degenerate configurations of the tri-component pairing $s + d_{x^2-y^2} e^{i\phi_1} + d_{xy} e^{i\phi_2}$.
 Symmetry operations which can generate the configuration from the one in (a$_1$)  are indicated on top of each figure,
 where $E$ is identity operation; $T$ is time reversal;   $C_6$, $C_3$, ..., $C_6^5$ are rotations; and $l_1$, $l_2$, ..., $l_6$ are reflections.
 In all panels, $\phi_1$ and $\phi_2$ are not perpendicular with each other.
In panels (a$_3$), (a$_5$), (b$_3$), and (b$_5$), $\psi_1$ and $\psi_s$ are not exactly collinear, and $\psi_2$ is not precisely equal to $\pm \pi/2$.
The parameters in Eq. \eqref{eq:F} are chosen as $\alpha_s=-N_F$, $\alpha_d=-3.179N_F$, $\beta_s=2.635N_F/T_c^2$, $\beta_d=0.790N_F/T_c^2$, $\gamma=0$, $g_{dd}=2.640N_F/T_c^2$, $g_{sd}=0.275N_F/T_c^2$, and $g'_{sd}=-1.525N_F/T_c^2$,
where $N_F$ is the density of states at the Fermi level and $T_c$ is the superconducting transition temperature.}
    \label{fig:TC6vdds}
\end{figure*}

Next we focus on the $f_1^\prime$ and $f_2^\prime$ terms in Eq. (\ref{eq:Fphi}). 
Since $g'_{sd}$ can be either positive or negative, the sign of $f'_1$ is determined by the product of $g'_{sd}$ and $\left( |\psi_1|^2-2|\psi_2|^2 \right)$, while the sign of $f'_2$ is determined by $g'_{sd}$.
When $g'_{sd}>0$, we have $f'_2<0$, then $2\phi_2 - \phi_1$ tends to take values of $2n\pi$.
In this case, if $|\psi_1|^2-2|\psi_2|^2>0$, then $\phi_1$ tends to be $(2m+1)\pi$, resulting in $\phi_2=(2m+2n+1)\pi/2$;
whereas if $|\psi_1|^2-2|\psi_2|^2<0$, $\phi_1$ tends to take value of $2m\pi$, resulting in $\phi_2=(m+n)\pi$,
where both $m$ and $n$ are integers.
Similar analysis can be performed for a negative $g'_{sd}$. 
The inclusion of the $g'_{sd}$ term makes the competition between $\phi_1$, $\phi_2$ and $\phi_2-\phi_1$ more complex, depending on the specific parameters taken in Eq. \eqref{eq:F}.
Notice that it is impossible for $\phi_1$ and $\phi_2$ to satisfy all the constraints set by $f_0$, $f_1$, $f_2$, $f_1^\prime$, and $f_2^\prime$. 

For a full treatment, in order to determine the pattern of the three order parameters, an iterative numerical method is applied to obtain the solution of the pairing gap function by minimizing Eq. \eqref{eq:F}.
The obtained results of pairing configurations are shown in Fig. \ref{fig:C6vdds}.
The parameters in free energy in Eq. \eqref{eq:F} to obtain Fig. \ref{fig:C6vdds} are chosen as $\alpha_s=-N_F$, $\alpha_d=-3.179N_F$, $\beta_s=2.635N_F/T_c^2$, $\beta_d=0.790N_F/T_c^2$, $g_{dd}=2.640N_F/T_c^2$, $g_{sd}=0.275N_F/T_c^2$, and $g'_{sd}=-1.525N_F/T_c^2$, where $N_F$ is the density of states at the Fermi level and $T_c$ is the superconducting transition temperature.
As previously discussed, the three parameters $g_{dd}$, $g_{sd}$, and $g_{sd}^\prime$ have significant impacts on the relative phases among different pairing components, thereby require careful consideration.
On the other hand, $\gamma$ is chosen to be 0 for simplification, as its value does not have a decisive influence on the relative phases and the symmetry breaking pattern. 
The obtained order parameters for this particular choice of parameters are $|\psi_s|=0.604k_BT_c$,  $|\psi_1|=1.029k_BT_c$,  $|\psi_2|=0.962k_BT_c$, $\phi_1=0.383\pi$, and $\phi_2=0.835\pi$. 
We note that the phase difference between the two $d$-wave components  is $\phi_2-\phi_1=0.452\pi$, which is not equal to $\pi/2$ as in the chiral $d+id$ case. 

The symmetry-breaking pattern of the configuration in Fig. \ref{fig:TC6vdds} (a$_1$) can be determined as
\begin{equation}
C_{6v} \times \mathbb{Z}_2^T \to C_2
\label{eq:symmetrybreaking}
\end{equation}
where $\mathbb{Z}_2^T$ is the $\mathbb{Z}_2$ group generated by time reversal operation.
Eq. (\ref{eq:symmetrybreaking}) is straightforward to be verified because the $\pi$-rotation around $z$-axis takes  $x,y$ to $-x,-y$, resulting in $d_{x^2-y^2} \to d_{x^2-y^2}$ and $d_{xy} \to d_{xy}$.
Except for $C_2$, all other symmetries are spontaneously broken in the ground state in Fig. \ref{fig:TC6vdds} (a$_1$).
Since $|C_{6v} \times \mathbb{Z}^T_2| / |C_2| = 12$, where $|...|$ represents the number of group elements,
there are 12 degenerate solutions of the ground state pairing configurations.
The other eleven degenerate configurations are shown in Fig. \ref{fig:TC6vdds} (a$_2$)-(a$_6$) and (b$_1$)-(b$_6$),
which can be obtained by performing the broken symmetry operations on the configuration in Fig. \ref{fig:TC6vdds} (a$_1$).
The symmetry operations that can be used to generate the corresponding configuration from Fig. \ref{fig:TC6vdds} (a$_1$) are indicated on top of each subfigure in Fig. \ref{fig:TC6vdds}.

\begin{table*}[t]
\centering
\begin{tabular}{|c|c|}
    \hline
    Symmetry Groups & G-L Free Energy \\
    \hline
    $C_{3v}$, $D_3$, $D_{3h}$, $D_{3d}$, $C_{6v}$,  $D_{6}$, $D_{6h}$ & $F_s^{(0)}+F_d^{(0)}+F^{(4)}$ \\
    \hline
    $D_{4d}$ & $F_s^{\prime(0)}+F_d^{\prime(0)}+F^{\prime(4)}$ \\
    \hline
\end{tabular}
\captionsetup{justification=raggedright}
\caption{Point groups with $d_1$-$d_2$ degeneracy and the corresponding free energy.
The explicit forms of $F_s^{(0)}$, $F_d^{(0)}$, $F^{(4)}$, $F_s^{'(0)}$, $F_d^{'(0)}$, and $F^{'(4)}$ are given in Eqs. (\ref{eq:F0},\ref{eq:F4},\ref{eq:F2prime},\ref{eq:F4prime}). 
}
\label{tab:groupF}
\end{table*}


\subsection{Other lattice symmetries}

On the free energy level, degenerate $d_{x^2-y^2}$ and $d_{xy}$ pairings can occur for other lattice symmetries as well, not just the $C_{6v}$ symmetry considered in Sec. \ref{subsec:free_energy_C6v}.
In this subsection, we discuss the general forms of free energies when there is a coexistence of $s$-, $d_{x^2-y^2}$-, and $d_{xy}$-pairing components, focusing on the special cases for planar point groups where $d_{x^2-y^2}$- and $d_{xy}$-channels are degenerate,
namely, they form a two-dimensional irreducible representation of the symmetry group. 

It turns out that there are eight planar point group symmetries that satisfy the condition of degenerate $d_{x^2-y^2}$ and $d_{xy}$ pairings,
including  $C_{3v}$, $D_3$, $D_{3h}$, $D_{3d}$, $C_{6v}$, $D_{6}$, $D_{6h}$, and $D_{4d}$.
Among the eight  point group symmetries, seven of them -- from $C_{3v}$ to $D_{6h}$ -- share the same form of free energy up to quartic order as the $C_{6v}$ case given in Eq. (\ref{eq:F}), whereas the  $D_{4d}$ case has a different form.
More explicitly, the free energy $F^\prime$ for $D_{4d}$ is given by
\begin{eqnarray}
F^\prime=F_{s}^{\prime(0)} + F_{d}^{\prime(0)}+F^{\prime(4)},
\label{eq:Fprime}
\end{eqnarray}
in which
\begin{eqnarray}
F_s^{\prime(0)}&=&\alpha_s^\prime|\psi_s|^2+\beta_s^\prime|\psi_s|^4,\nn\\
F_d^{\prime(0)}&=&\alpha_d^\prime \left( |\psi_1|^2+|\psi_2|^2 \right) + \beta_d^\prime  \left( |\psi_1|^2+|\psi_2|^2 \right)^2,
\label{eq:F2prime}
\end{eqnarray}
and 
\begin{eqnarray}
F^{\prime(4)}&=&\gamma_{dd}|\psi_1|^2|\psi_2|^2+\gamma_{sd}|\psi_s|^2 \left( |\psi_1|^2+|\psi_2|^2 \right) \nn\\
&+& g_{dd}^\prime \left(\psi_1^2 \psi_2^{*2}+\psi_1^{*2} \psi_2^2 \right)\nn\\
&+& g_{sd}^\prime \left[ \psi_s^{*2} \left( \psi_1^2 + \psi_2^2 \right) + \psi_s^2 \left( \psi_1^{*2} + \psi_2^{*2} \right)\right].
\label{eq:F4prime}
\end{eqnarray}
In this case, we also expect a mixture of $d_{x^2-y^2}$-, $d_{xy}$-, and $s$-wave pairing symmetries when  $g^\prime_{dd}>0$ and $g^\prime_{sd}>0$ in Eq. (\ref{eq:F4prime}).


\section{Topological chiral pairing}
\label{sec:topo}

\subsection{Microscopic model}

Now we turn to a microscopic model.
Retaining terms up to the nearest neighbors, it is direct to construct the Bogoliubov–de Gennes (BdG) Hamiltonian for the tri-component superconducting pairing on a two-dimensional honeycomb lattice.
The BdG Hamiltonian can be written as
\begin{equation}
H=\sum_{\boldsymbol{k}} \Psi_{\boldsymbol{k}}^{\dagger} h_{\boldsymbol{k}} \Psi_{\boldsymbol{k}},
\end{equation}
in which $\Psi_{\boldsymbol{k}} = (c_{\boldsymbol{k}_A,\uparrow}~c_{\boldsymbol{k}_B,\uparrow}~c_{-\boldsymbol{k}_A,\downarrow}^\dagger~c_{-\boldsymbol{k}_B,\downarrow}^\dagger)^T$,
where $c_{\boldsymbol{k}_j,\sigma}^\dagger$ and $c_{\boldsymbol{k}_j,\sigma}$ represent the creation and annihilation operators, respectively, for an electron with momentum $\boldsymbol{k} = (k_x, k_y)$ and spin $\sigma$ in the sublattice  $j=A,B$; 
$h_{\boldsymbol{k}}$ is the $4\times4$ BdG Hamiltonian matirx containing the normal-state Hamiltonian $H_0(\boldsymbol{k})$ and the pairing term $\Delta(\boldsymbol{k})$, given by
\begin{equation}
h_{\boldsymbol{k}}=
 \begin{pmatrix}
H_0(\boldsymbol{k}) & \Delta(\boldsymbol{k}) \\
\Delta^\dagger(\boldsymbol{k}) & -H^T_0(-\boldsymbol{k})
\end{pmatrix}.
\label{eq:hk1}
\end{equation}

In the absence of a staggered potential \cite{Cao2013,Hadad2016,Qi2023}, the normal-state Hamiltonian can be expressed as
\begin{equation}
H_0(\boldsymbol{k})=\epsilon_x(\boldsymbol{k}) \sigma_x + \epsilon_y(\boldsymbol{k}) \sigma_y - \mu \sigma_0,
\end{equation}
where
\begin{eqnarray}
\epsilon_x(\boldsymbol{k})&=&-t \sum_{i=1}^3 \cos{(\boldsymbol{k} \cdot \boldsymbol{a_i})}, \notag\\
\epsilon_y(\boldsymbol{k})&=&t \sum_{i=1}^3 \sin{(\boldsymbol{k} \cdot \boldsymbol{a_i})},
\end{eqnarray}
in which $\sigma_\alpha$ ($\alpha=x,y,z$) are $2 \times 2$ Pauli matrices that encode the sublattice degree of freedom,
$\sigma_0$ is the $2 \times 2$ identity matrix, $\mu$ is the chemical potential, $t$ is the nearest-neighbor hopping amplitude,
and $\boldsymbol{a_i} ~ (i=1,2,3)$ are vectors  defined in Fig. \ref{fig:nearestn}.
The Hamiltonian $H_0(\boldsymbol{k})$ describes the kinetic energy and nearest-neighbour hopping of electrons in the non-superconducting state of the system.

Considering the superconducting pairing in the chiral spin-singlet state, the tri-component pairing term can be expressed as
\begin{equation}
\Delta(\boldsymbol{k})=\Delta_s(\boldsymbol{k})+\Delta_{x^2-y^2}(\boldsymbol{k})e^{i\phi_1}+\Delta_{xy}(\boldsymbol{k})e^{i\phi_2},
\label{eq:Deltak0}
\end{equation}
where
\begin{equation}
\Delta_s(\boldsymbol{k})=|\psi_s| \sigma_0, 
\label{eq:Delta_s}
\end{equation}
and
\begin{flalign}
&\Delta_{x^2-y^2}(\boldsymbol{k})=\notag\\
&|\psi_1| \left \{ \left[ \cos{(k_xa)} - \cos{\left( \frac{1}{2}k_xa \right)} \cos{\left( \frac{\sqrt{3}}{2}k_ya \right)} \right]\sigma_x \right. \notag\\
& \left. -\left[ \sin{(k_xa)} + \sin{\left( \frac{1}{2}k_xa \right)} \cos{\left( \frac{\sqrt{3}}{2}k_ya \right)} \right] \sigma_y \right \}, \notag\\
&\Delta_{xy}(\boldsymbol{k})=\notag\\
&\sqrt{3}|\psi_2| \left \{ \left[ - \sin{\left( \frac{1}{2}k_xa \right)} \sin{\left( \frac{\sqrt{3}}{2}k_ya \right)} \right]\sigma_x \right. \notag\\
& \left. +\left[ \cos{\left( \frac{1}{2}k_xa \right)} \sin{\left( \frac{\sqrt{3}}{2}k_ya \right)} \right] \sigma_y \right \}.
\label{eq:Deltas12}
\end{flalign}
According to the calculations in Appendix \ref{app:dwave}, $(\Delta_{x^2-y^2}, \Delta_{xy})$ transform in the same way as $(d_{x^2-y^2}, d_{xy})$ under the $C_{6v}$ group.
Hence $(\Delta_{x^2-y^2}, \Delta_{xy})$ represent $d$-wave pairings for the discrete symmetry group $C_{6v}$.

Substituting Eqs. (\ref{eq:Delta_s},\ref{eq:Deltas12}) into Eq. \eqref{eq:Deltak0}, we obtain
\begin{equation}
\Delta(\boldsymbol{k})=\Delta_s \sigma_0 + \Delta_x(\boldsymbol{k}) \sigma_x + \Delta_y(\boldsymbol{k}) \sigma_y,
\label{eq:Deltak1}
\end{equation}
where
\begin{equation}
\Delta_s=|\psi_s|, 
\end{equation}
and
\begin{flalign}
&\Delta_x(\boldsymbol{k})=\notag\\
&|\psi_1|e^{i\phi_1} \left[ \cos{(k_xa)} - \cos{\left( \frac{1}{2}k_xa \right)} \cos{\left( \frac{\sqrt{3}}{2}k_ya \right)} \right] \notag\\
& -\sqrt{3}|\psi_2|e^{i\phi_2} \left[ \sin{\left( \frac{1}{2}k_xa \right)} \sin{\left( \frac{\sqrt{3}}{2}k_ya \right)} \right], \notag\\
&\Delta_y(\boldsymbol{k})=\notag\\
&-|\psi_1|e^{i\phi_1} \left[ \sin{(k_xa)}+\sin{\left( \frac{1}{2}k_xa \right)} \cos{\left( \frac{\sqrt{3}}{2}k_ya \right)} \right] \notag\\
& +\sqrt{3}|\psi_2|e^{i\phi_2} \left[ \cos{\left( \frac{1}{2}k_xa \right)} \sin{\left( \frac{\sqrt{3}}{2}k_ya \right)} \right].
\end{flalign}
The $4\times4$ BdG Hamiltonian can be expressed as
\begin{eqnarray}
h_{\boldsymbol{k}} &=& \left[ \text{Re} (\Delta_x) \sigma_x + \text{Re} (\Delta_y) \sigma_y + \Delta_s \sigma_0 \right] \tau_x \notag\\
&&- \left[ \text{Im} (\Delta_x) \sigma_x + \text{Im} (\Delta_y) \sigma_y \right] \tau_y \notag\\
&&+ \left( \epsilon_x \sigma_x + \epsilon_y \sigma_y -\mu \sigma_0 \right) \tau_z,
\label{eq:BdGhk}
\end{eqnarray}
where $\tau_\alpha$ ($\alpha=x,y,z$) are the Pauli matrices in the particle-hole channel.
Diagonalization of $h_{\boldsymbol{k}}$ gives two pairs of energy eigenvalues $\pm E_i(\boldsymbol{k})$ for each momentum $\boldsymbol{k}$,
\begin{flalign}
&E_i(\boldsymbol{k})=\notag\\
&\sqrt{|\Delta_x|^2+|\Delta_y|^2+|\Delta_s|^2+\epsilon_x^2+\epsilon_y^2+\mu^2+(-)^i D_{\boldsymbol{k}}},
\end{flalign}
where $i=1,2$ and
\begin{align}
D_{\boldsymbol{k}}^2 
&= 4 \mu^2 (\epsilon_x^2 + \epsilon_y^2) + 2|\Delta_x|^2 |\Delta_y|^2 - \left( \Delta_x^2 \Delta_y^{*2} + \Delta_y^2 \Delta_x^{*2} \right) \notag\\
&+ 4 \left[\epsilon_x^2 |\Delta_y|^2 + \epsilon_y^2 |\Delta_x|^2 - \epsilon_x \epsilon_y \left( \Delta_x  \Delta_y^*+\Delta_y \Delta_x^* \right) \right] \notag\\
&+ \Delta_s^2 \left[ \left(\Delta_x+\Delta_x^* \right)^2 + \left( \Delta_y+\Delta_y^* \right)^2 \right] \notag\\
&- 8 \mu \Delta_s \left[ \epsilon_x \text{Re} (\Delta_x) + \epsilon_y \text{Re} (\Delta_y) \right].
\end{align}

\begin{figure}[ht]
    \centering
    \begin{subfigure}[b]{0.4\textwidth}
        \centering
        \includegraphics[width=\textwidth]{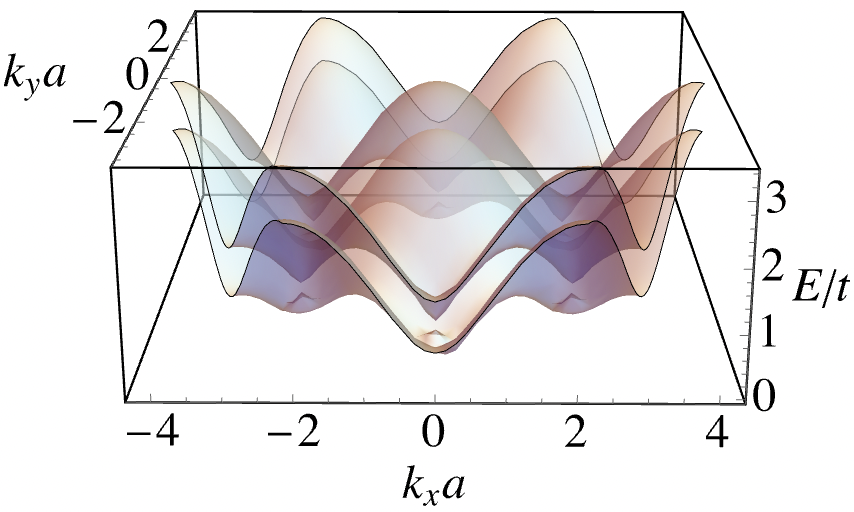}
        \caption{}
        \label{fig:sub1}
    \end{subfigure}
    \hfill
    \begin{subfigure}[b]{0.295\textwidth}
        \centering
        \includegraphics[width=\textwidth]{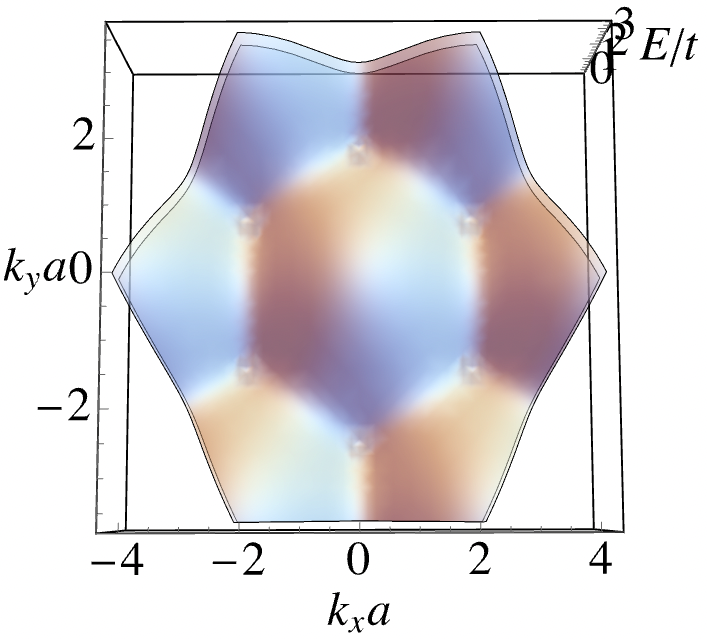}
        \caption{}
        \label{fig:sub2}
    \end{subfigure}
    \hfill
    \begin{subfigure}[b]{0.18\textwidth}
        \centering
        \includegraphics[width=\textwidth]{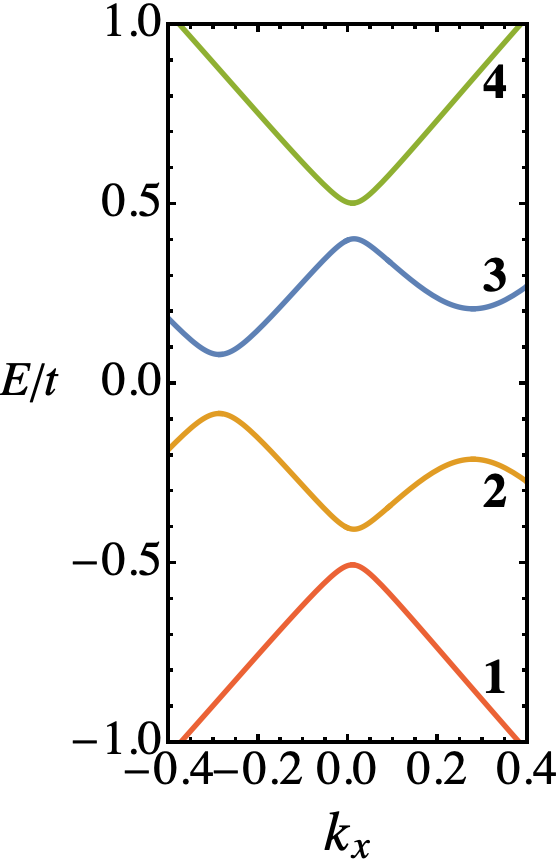}
        \caption{}
        \label{fig:sub3}
    \end{subfigure}
    \captionsetup{justification=raggedright}
    \caption{Eigenvalues of the BdG Hamiltonian in Eq. \eqref{eq:BdGhk}.
Panel (a) presents three-dimensional plots of the positive $E_1$ and $E_2$ as functions of $\boldsymbol{k}$, where a gap opens at the Dirac points.
Panel (b) provides a top view of Panel (a).
Panel (c) illustrates the variation of the four eigenvalues with respect to $k_x$, with the $k_y$ coordinate fixed to be the $y$-component of the  $K$ point.
In all plots, we set $|\psi_s|=0.0604t$, $|\psi_1|=0.1029t$, $|\psi_2|=0.0962t$, $\phi_1=0.383\pi$, $\phi_2=0.835\pi$, and $\mu=-0.4t$.}
    \label{fig:Egap}
\end{figure}

The plots of $E_1(\boldsymbol{k})$ and $E_2(\boldsymbol{k})$ are depicted in Fig. \ref{fig:Egap} by taking $|\psi_s|=0.0604t$, $|\psi_1|=0.1029t$, $|\psi_2|=0.0962t$, $\phi_1=0.383\pi$, $\phi_2=0.835\pi$, and $\mu=-0.4t$,
which give a fully gapped energy spectrum. 
To more clearly demonstrate the gap, Fig. \ref{fig:Egap}(c) is created by fixing $k_y=4\pi/\left(3\sqrt{3}a\right)$ (which is the $y$-coordinate of the $K$ point) and plotting the curves of $\pm E_{1,2}(\boldsymbol{k})$ as functions of $k_x$ only.
It is evident that a gap exists at the Dirac point $K=(0,\frac{4\pi}{3\sqrt{3}a} )$, where $E \approx \pm0.454t$ and the gap size $\delta E \approx 0.103t$.


\subsection{Chern number}

A non-vanishing Chern number is the signature of a non-trivial topological property for two-dimensional systems, which is defined as \cite{Thouless1982,Simons1983}
\begin{equation}
C=\frac{1}{2i\pi}\int d^2k F_{xy}(\boldsymbol{k}),
\label{eq:Cn}
\end{equation}
in which the Berry connection $A_\alpha(\boldsymbol{k}) (\alpha=x,y)$ and the associated field strength $F_{xy}(\boldsymbol{k})$ are given by
\begin{eqnarray}
A_\alpha(\boldsymbol{k})&=&\langle n(\boldsymbol{k})|\partial_\alpha|n(\boldsymbol{k}) \rangle, \notag\\
F_{xy}(\boldsymbol{k})&=&\partial_xA_y(\boldsymbol{k})-\partial_yA_x(\boldsymbol{k}),
\label{eq:Fxy}
\end{eqnarray}
where $|n(\boldsymbol{k}) \rangle$ is a  normalized wave function of the $n$th energy band.
With the opening of the mass gap, the four eigenstates of the BdG Hamiltonian in Eq. \eqref{eq:BdGhk} are everywhere nondegenerate, so that a Chern number $C_n$ can be defined for each band $n$ as labeled in Fig. \ref{fig:Egap}(c).
At $T=0$K, the Chern numbers for corresponding energy bands are calculated as $C_1 = 1$, $C_2 = -3$, $C_3 = 3$, and $C_4 = -1$,
using the numerical method proposed in Ref. \onlinecite{Fukui2005,Kudo2019}.
The choices of order parameters and chemical potential are the same as in Fig. \ref{fig:Egap}, corresponding to Fig. \ref{fig:TC6vdds} (a$_1$). 
The sum of the Chern numbers of the two occupied bands (i.e., $n = 1, 2$) is $-2$, which is consistent with the characteristics of a chiral $d$-wave superconductivity.

\begin{figure}[h]
\centering
\includegraphics[width=0.49\textwidth]{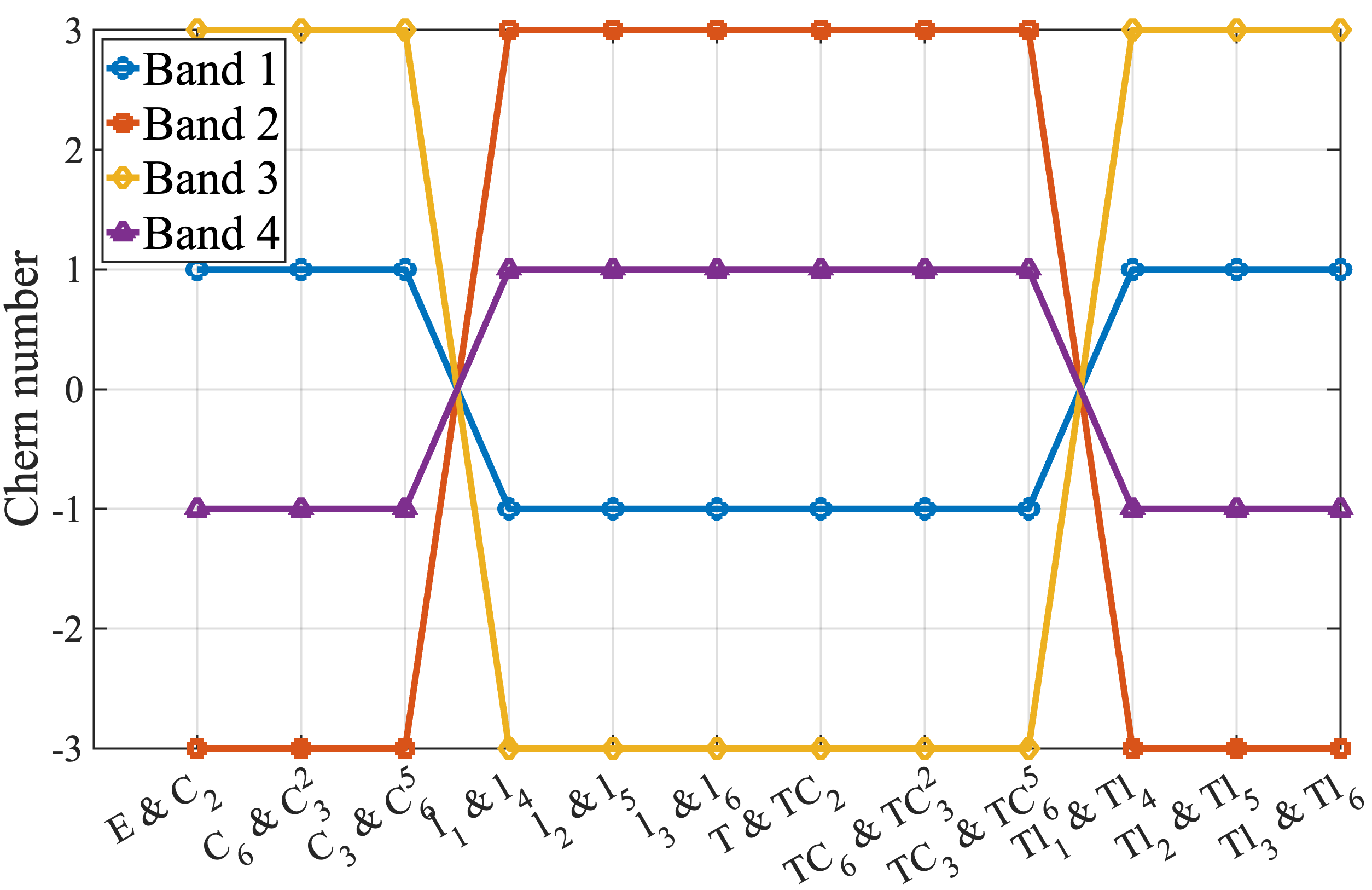}
 \captionsetup{justification=raggedright}
\caption{Evolution of Chern numbers under  symmetry operations in $C_{6v} \times \mathbb{Z}_2^T$.}
\label{fig:chern}
\end{figure}

The behavior of the Chern numbers under the transformations of the $C_{6v} \times \mathbb{Z}^T_2$ group elements is illustrated in Fig. \ref{fig:chern}.
If $\phi_2-\phi_1<\pi$, the Chern numbers are given by $C_1 = 1$, $C_2 = -3$, $C_3 = 3$, $C_4 = -1$, and the sum of the Chern numbers of the bands with negative energies is $-2$;
conversely, if $\phi_2-\phi_1>\pi$, the Chern numbers change sign, and the sum is $2$.
It can be observed that rotational transformations do not alter the Chern numbers, while mirror reflection and time-reversal transformations reverse the sign of the Chern numbers.

\subsection{Edge states}

When Chern number is nonzero, it is expected that chiral edge states emerge on the boundaries of the system \cite{Delplace2011,Obana2019}, 
which, for the case of topological superconductors,  are Majorana modes propagating uni-directionally along the sample boundaries \cite{Plotnik2014,MacDonald1990,Hatsugai1993,Sarma2015,Tanaka2024,Sun2017,Flensberg2021,Mercado2022}. 

\begin{figure}
\includegraphics[width=0.45 \textwidth]{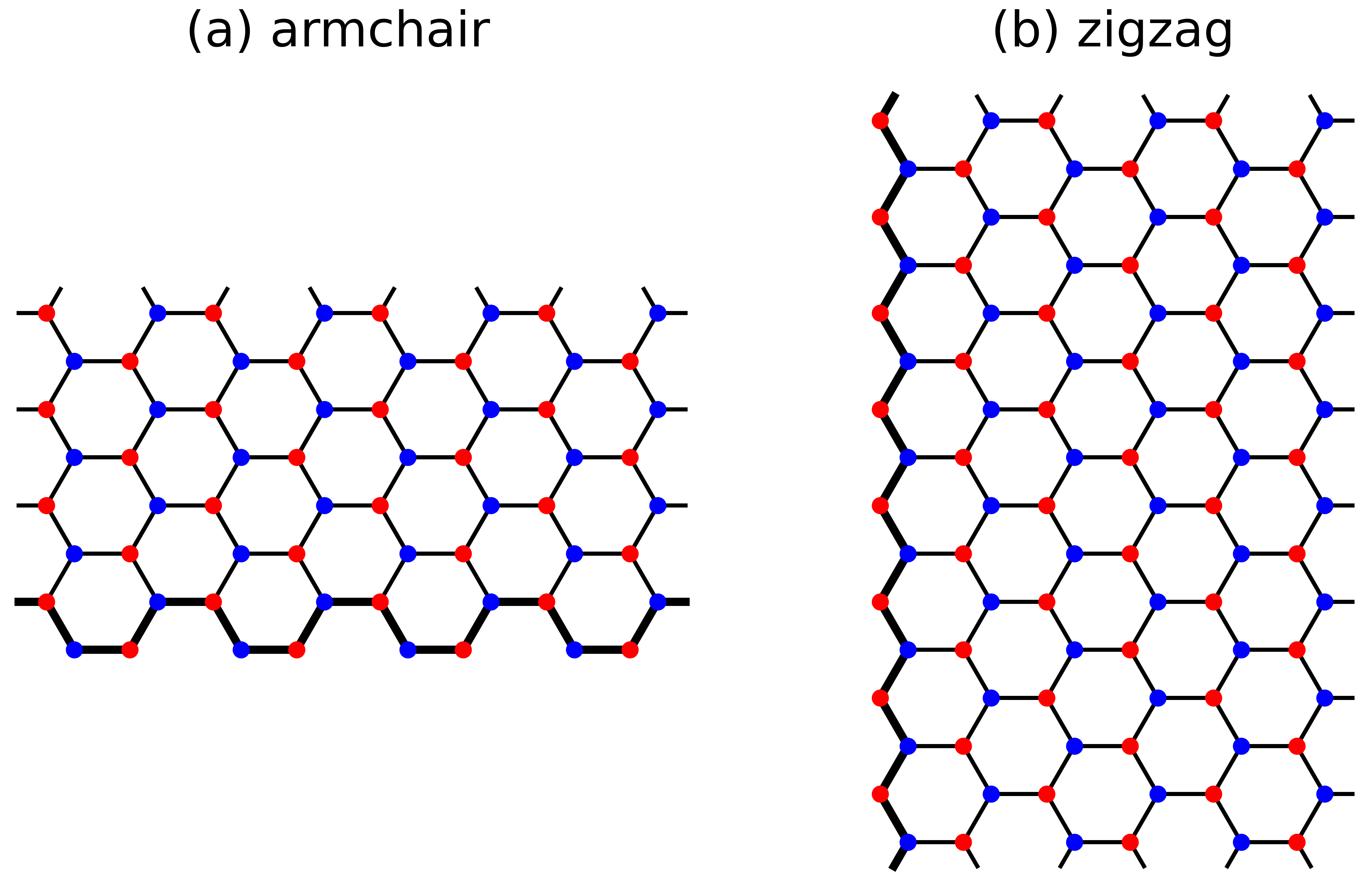}
\captionsetup{justification=raggedright}
\caption{Plots for (a) armchair and (b) zigzag edges on the honeycomb lattice,
in which the thickened bonds represent the bonds on the edges. 
}
\label{fig:armchair}
\end{figure}

\begin{figure}[ht]
    \centering
    \begin{subfigure}[b]{0.235\textwidth}
        \centering
        \includegraphics[width=\textwidth]{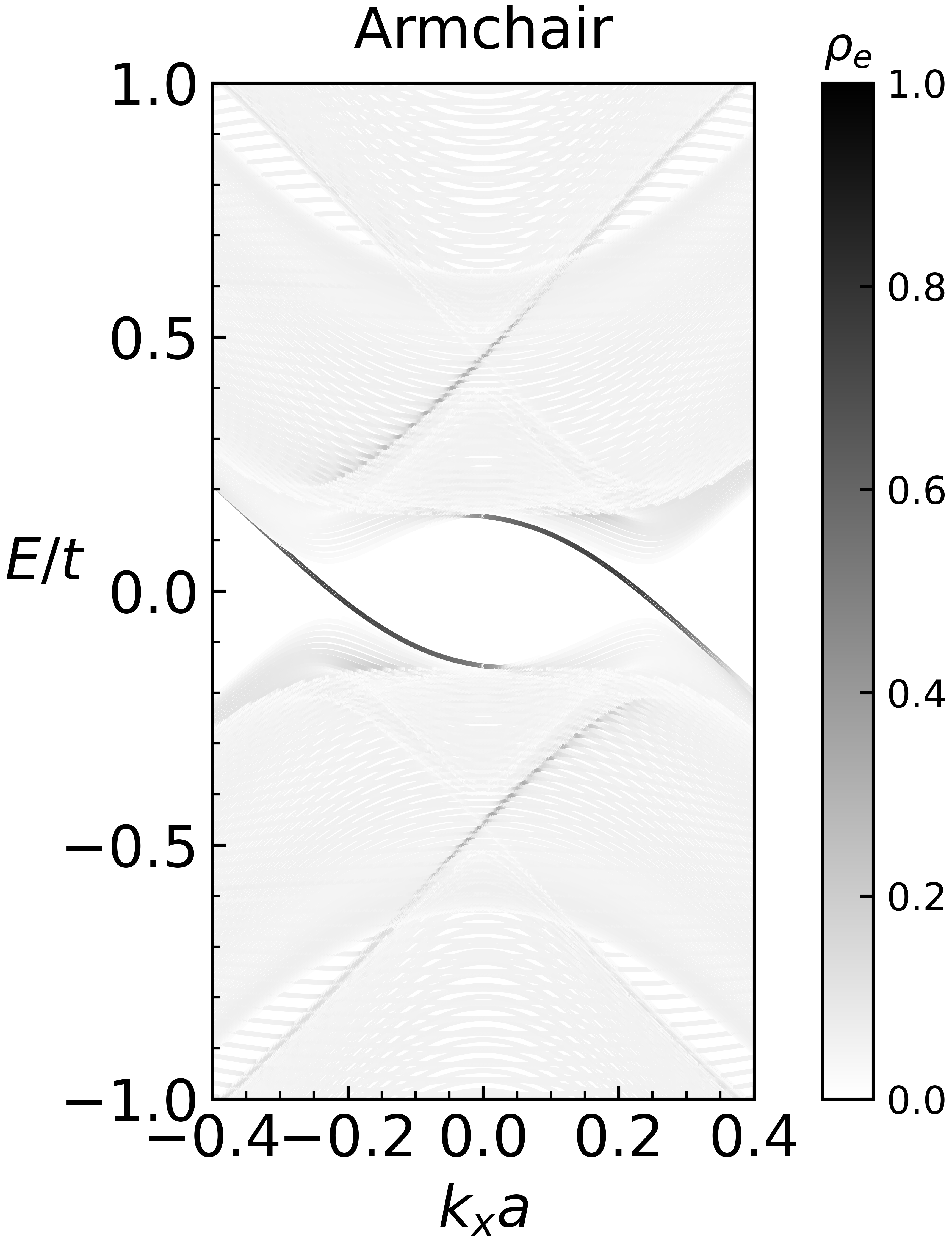}
        \caption{}
        \label{fig:sub1}
    \end{subfigure}
    \hfill
    \begin{subfigure}[b]{0.235\textwidth}
        \centering
        \includegraphics[width=\textwidth]{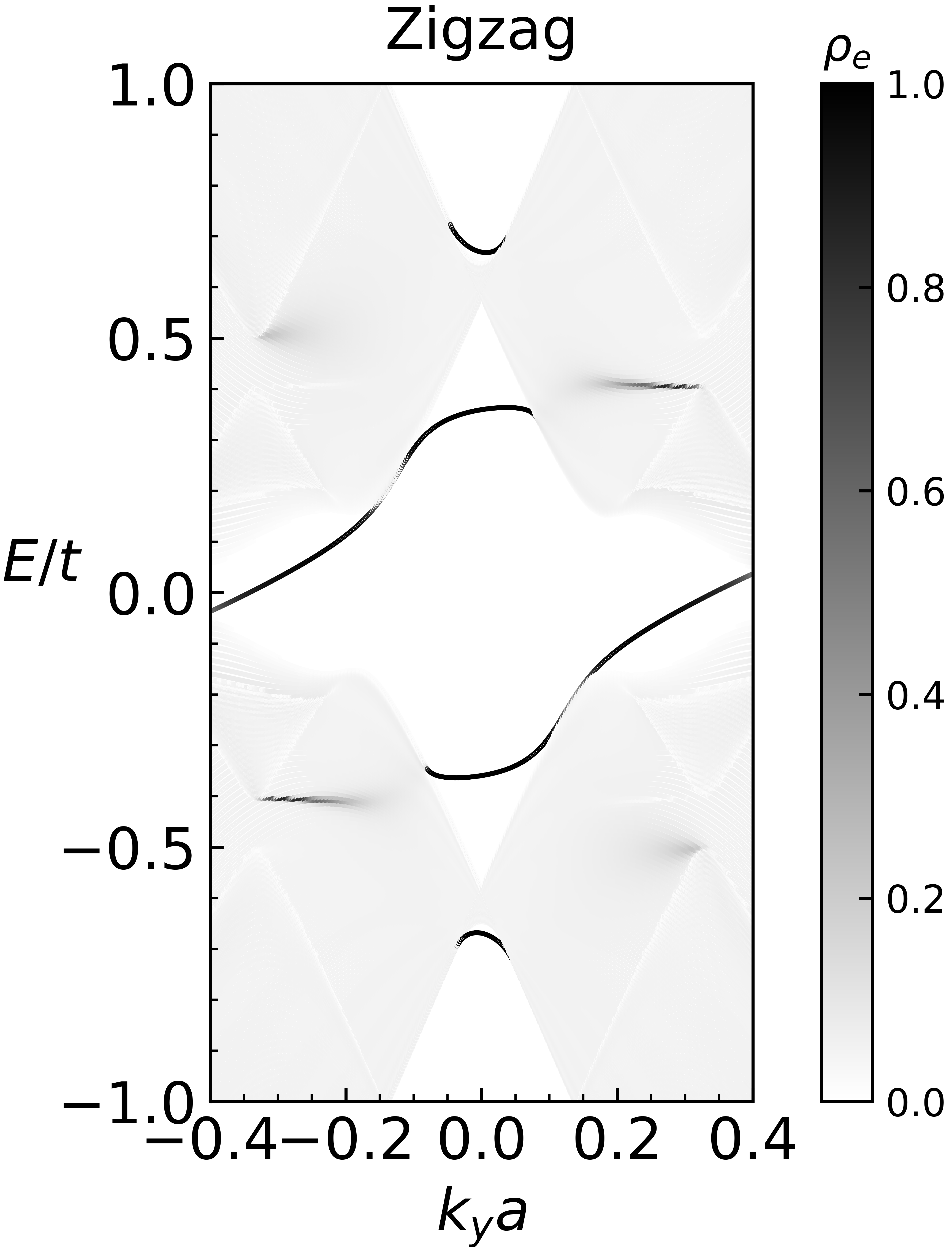}
        \caption{}
        \label{fig:sub2}
    \end{subfigure}
    \captionsetup{justification=raggedright}
    \caption{Energy spectrum for (a) armchair edge and (b) zigzag edge.
$\rho_e$ the integrated  wavefunction probabilities in the first ten stripes of unit cells close to the boundaries. 
The choices of order parameters and chemical potential are the same as in Fig. \ref{fig:Egap}.}
    \label{fig:edge}
\end{figure}

To further study the topological properties of the obtained tri-component superconducting pairing $s + d_{x^2-y^2} e^{i\phi_1} + d_{xy} e^{i\phi_2}$, we solve edge states for different boundary geometries, namely, armchair and zigzag as shown in Fig. \ref{fig:armchair},
by taking the 1D edge direction of the two-dimensional honeycomb lattice to be periodic and the perpendicular direction to be open and finite with 100 unit cells of the honeycomb lattice.
Fig. \ref{fig:edge} (a) and (b) display the numerical results for the energy spectrum for armchair and zigzag boundaries, respectively,
in which the horizontal axes are the momenta in the periodic directions, namely, $k_x$ for armchair and $k_y$ for zigzag (for definition of $x$- and $y$-directions, see Fig. \ref{fig:nearestn}),
and the vertical axis is the exictation energy $E$ in units of hopping $t$. 
The grey scales $\rho_e$ in Fig. \ref{fig:edge} (a,b) are defined as
\begin{equation}
\rho_e=\sum_{j=1}^{10}\sum_{\lambda=A,B} \sum_{\sigma=\uparrow,\downarrow}{\left|\psi_\sigma(j,\lambda)\right|}^2,
\end{equation}
in which 
$j$ is the index of the stripe of unit cells measured from the edge in the direction perpendicular to the edge,
$\lambda$ is the index for  sublattice sites within a unit cell,
$\sigma$ is the spin index,
and $\psi_\sigma(j,\lambda)$ is the wavefunction for the corresponding eigenstate. 
Notice that $\rho_e$ represents the integrated  wavefunction probabilities in the first ten stripes of unit cells close to the boundaries,
thereby can be used for calibrating the degree of localization of the wavefunctions near the boundaries. 

As can be seen from Fig. \ref{fig:edge}(a),  two well-defined dispersive mid-gap modes can be clearly observed for the armchair case.  
The relatively dark colors of the two mid-gap lines undoubtedly hint their edge state nature.  
Moreover, the two chiral edge modes both cross $E=0$ in Fig. \ref{fig:edge}(a), consistent with a total Chern number of the two negative energy bands being equal to $\pm 2$.
As for the case of zigzag edge shown in Fig. \ref{fig:edge}(b), two chiral edge modes crossing $E=0$ can also be observed, again consistent with the bulk Chern number. 

\section{Anisotropic AC Hall conductivity}
\label{sec:Hall}

Under the influence of incident light, the refractive index of a material changes,
which in turn alters the polarization state of the light.
This is known as the optical Kerr effect \cite{Argyres1955,Bree2011}, and the angle of polarization change is called the Kerr angle, which  can be used  as a signal for chiral topological phase \cite{Okada2016,Ovalle2024}.
In the optical Kerr effect, we focus on the interaction between light and matter,
which is influenced by the anomalous AC Hall conductivity $\sigma_H(\omega)$ \cite{Hu2022,Parent2020,Taylor2012}.
The Kerr angle, $\theta_K$, is directly related to $\sigma_H(\omega)$,
and as a result, the presence of $\sigma_H(\omega)$ can be used as an experimental signature for the existence of time-reversal symmetry breaking.

\subsection{Expression for AC Hall conductivity}

The anomalous AC Hall conductivity is the antisymmetric part of the optical Hall conductivity, 
\begin{equation}
\sigma_H(\omega)=\frac{1}{2}\lim_{\boldsymbol{q} \to 0} \left[ \sigma_{xy}(\boldsymbol{q},\omega)-\sigma_{yx}(\boldsymbol{q},\omega) \right],
\label{eq:sigma}
\end{equation}
in which $\omega$ is the frequency of the incident light, and the optical Hall conductivity $\sigma_{xy}(\boldsymbol{q},w)$ is related to the current-current correlator $\pi_{xy}(\boldsymbol{q},\omega)$ via
\begin{equation}
\sigma_{xy}(\boldsymbol{q},\omega)=\frac{1}{\hbar \omega} \pi_{xy}(\boldsymbol{q},\omega).
\end{equation}
The current-current correlator $\pi_{xy}(\boldsymbol{q},\omega)$ is defined as
\begin{equation}
\pi_{xy}(\boldsymbol{q},\omega)=\int_0^\infty dte^{i\omega t}\left \langle \left[ \hat{J}_x^\dagger(\boldsymbol{q},t), \hat{J}_y(\boldsymbol{q},0) \right] \right \rangle,
\label{eq:pixy}
\end{equation}
in which $\hat{J}_\alpha=e\sum_{\boldsymbol{k}} \Psi_{\boldsymbol{k}}^\dagger \hat{v}_\alpha \Psi_{\boldsymbol{k}}$ is the $\alpha$'th component ($\alpha=x,y$) of the current operator, 
and $\hat{v}_\alpha$ is the $\alpha$'th component of the velocity operator in Nambu notation given by
\begin{equation}
\hat{v}_\alpha= \frac{1}{\hbar} \left( \sigma_z \otimes \sigma_0 \right) \partial_{k_\alpha} h_{\boldsymbol{k}}^0,
\end{equation}
where $h_{\boldsymbol{k}}^0$ is the normal part of the BdG Hamiltonian in Eq. \eqref{eq:BdGhk}, namely,
\begin{equation}
h_{\boldsymbol{k}}^0 = \left( -\mu \sigma_0 + \epsilon_x \sigma_x + \epsilon_y \sigma_y \right) \tau_z.
\label{eq:hk0}
\end{equation}

Under time-reversal transformation, the current correlator changes as $\pi_{xy} \to \pi_{yx}$;
and under mirror reflection transformation along $\hat{x}$- or $\hat{y}$-axis, the current correlator changes as $\pi_{xy} \to -\pi_{xy}$ (for details, see Appendix \ref{app:mirror}),
both of which result in a sign change of $\sigma_H$.
Therefore, if either of the two above symmetries is unbroken, $\sigma_H$ must be zero in order to preserve the invariance of the observable quantity.
Hence, to obtain a nonzero AC Hall conductivity, both time-reversal and mirror symmetries along the $\hat{x}$- and $\hat{y}$-axes must be broken.

\begin{figure*}[ht]
    \centering
    \begin{subfigure}[b]{0.329\textwidth}
        \centering
        \includegraphics[width=\textwidth]{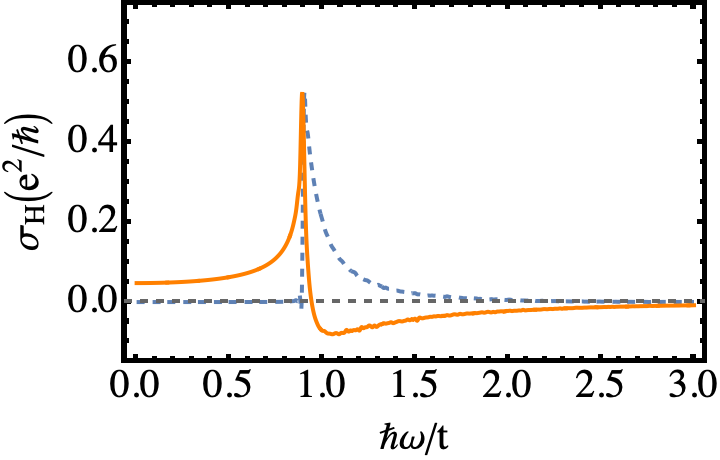}
        \caption{}
        \label{fig:sigmaHE}
    \end{subfigure}
    \hfill
    \begin{subfigure}[b]{0.329\textwidth}
        \centering
        \includegraphics[width=\textwidth]{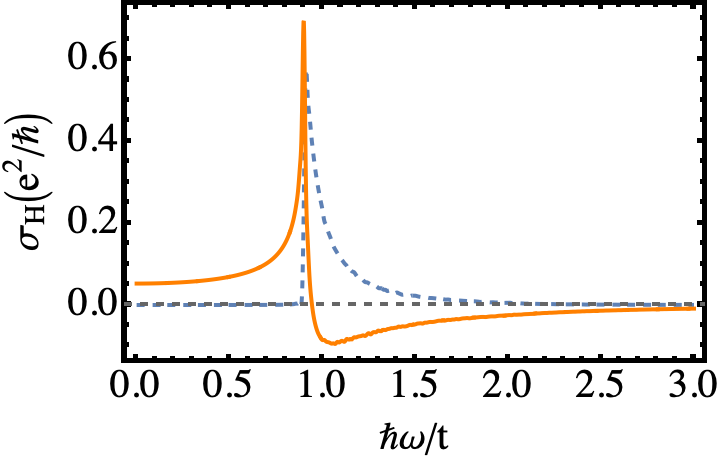}
        \caption{}
        \label{fig:sigmaHC6}
    \end{subfigure}
    \hfill
    \begin{subfigure}[b]{0.329\textwidth}
        \centering
        \includegraphics[width=\textwidth]{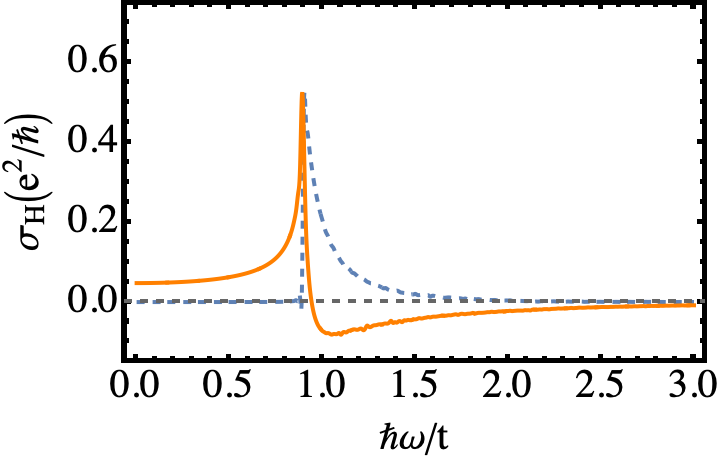}
        \caption{}
        \label{fig:sigmaHC3}
    \end{subfigure}
   \captionsetup{justification=raggedright}
\caption{
Anisotropy of the real (solid line) and imaginary (dashed line) parts of the anomalous AC Hall conductivity in Eq. \eqref{eq:sigmaH}.
Panel (a) shows the line-shape for $\sigma_H(\omega)$ without any symmetry operations,
Panel (b) and (c) present the line-shapes for $\sigma_H(\omega)$ when the coordinate axes are rotated by $\pi/3$ and $2\pi/3$ angles around the $z$-axis, respectively.
Both real part and imaginary part in all panels reach a peak at $\hbar \omega \approx 0.9t$. 
The fact that $\sigma_H \neq 0$ demonstrates that the system is in a chiral topological superconducting phase,
and the change in $\sigma_H$ after rotation in Panel (b) and (c) reflects the anisotropy of the anomalous AC Hall conductivity under the tri-component pairing configuration.
In this figure, we use the same parameters as in Fig. \ref{fig:Egap}, and set the temperature $k_BT=0.01t$.}
\label{fig:sigmaH}
\end{figure*}

Specifically, in superconducting systems with a multi-band feature and a symmetry-breaking pattern of $C_{6v} \times \mathbb{Z}^T_2 \to C_2$ as described by Eq. \eqref{eq:BdGhk}, the condition for generating a nonzero $\sigma_H$ is fulfilled. 
A straightforward evaluation from Eq. (\ref{eq:sigma}--\ref{eq:hk0}) yields the anomalous AC Hall conductivity as follow,
\begin{widetext}
\begin{equation}
\sigma_H(\omega)=\lim_{i\nu_m \to \omega+i\epsilon} \frac{e^2}{\hbar\beta} \int \frac{d^2k}{(2\pi)^2} \sum_{\omega_n}
\frac{2\hbar^3(\nu_m+2\omega_n)^2\left( v_x^*v_y-v_xv_y^* \right) \left[ \mu \left( \Delta_x^* \Delta_y - \Delta_x \Delta_y^* \right) + 2i\Delta_s \left( \epsilon_x \text{Im} \Delta_y -\epsilon_y\text{Im} \Delta_x \right) \right]}{\left( \hbar^2 \omega_n^2+E_1^2 \right) \left( \hbar^2 \omega_n^2+E_2^2 \right) \left[ \hbar^2 (\omega_n+\nu_m)^2+E_1^2 \right] \left[ \hbar^2 (\omega_n+\nu_m)^2+E_2^2 \right]}.
\label{eq:sigmaH}
\end{equation}
\end{widetext}
where $\nu_m$ is the bosonic Matsubara frequency, $\omega_n$ is the fermionic Matsubara frequency, $\beta=1/k_BT$,
$v_\alpha= (1/ \hbar) \partial_{k_\alpha} \left( \epsilon_x-i\epsilon_y \right)$, $\alpha=x,y$,
and $\epsilon$ represents a positive infinitesimal here.
For the tri-component pairing state $s+d_{x^2-y^2}e^{i\phi_1}+d_{xy}e^{i\phi_2}$ with time-reversal symmetry breaking, the leading term of the vertex correction is zero and thus can be neglected \cite{Goryo2008}.
Detailed derivation of Eq. (\ref{eq:sigmaH}) is included in Appendix \ref{app:derive_Hall}.
The reason why $\sigma_H(\omega)$ is nonzero can be directly seen from Eq. (\ref{eq:sigmaH}) as follows. 
In order for the Hall signal to be non-vanishing, the integrand in Eq. \eqref{eq:sigmaH} must be even under the reflection $x \leftrightarrow y$.
Indeed, the eigen-energy $E_\alpha$ in the denominator remains unchanged under the exchange of the $x$ and $y$ indices,
whereas the terms involving $\left( v_x^*v_y-v_xv_y^* \right)$, $\left( \Delta_x \Delta_y^* - \Delta_x^* \Delta_y \right)$, and $\left( \epsilon_x \text{Im} \Delta_y -\epsilon_y\text{Im} \Delta_x \right)$ in the numerator  change sign under $x \leftrightarrow y$, so that the overall numerator is also even.
As a result, Eq. \eqref{eq:sigmaH} satisfies the condition for generating a non-vanishing AC Hall conductivity.

\subsection{Numerical results for AC Hall conductivity}

The line-shapes for both the real and imaginary parts of $\sigma_{H}(\omega)$ as functions of $\omega$ are plotted in Fig. \ref{fig:sigmaH} (a), where the same parameters for the chemical potential and superconducting pairing are used as in Fig. \ref{fig:Egap},
and the temperature is set to be $k_BT=0.01t$.
From Fig. \ref{fig:sigmaH} (a), it can be observed that when the incident light energy satisfies $\hbar \omega = \left[ E_1(\boldsymbol{k})+E_2(\boldsymbol{k}) \right]_{\mathrm{min}} \approx 0.9t$, 
both $\text{Re}(\sigma_H)$ and $\text{Im}(\sigma_H)$ exhibit peaks.
The peak position in $\text{Im}(\sigma_H)$ can be understood from resonances. 
Notice that $\text{Im}(\sigma_H)$ contains delta-functions $\delta(E_1+E_2+\hbar \omega)$ and $\delta(E_1+E_2-\hbar \omega)$ (see Appendix \ref{app:derive_Hall} for details). 
The energy conservation constraint in $\delta(E_1+E_2+\hbar \omega)$ cannot be satisfied for positive $\omega$,
and the constraint in $\delta(E_1+E_2-\hbar \omega)$ can be satisfied only when $\omega$ is above the two-particle continuum, i.e., $\hbar\omega\geq \left[ E_1(\boldsymbol{k})+E_2(\boldsymbol{k}) \right]_{\mathrm{min}}$.
This is the reason for the onset of a nonzero $\text{Im}(\sigma_H)$ at $\left[ E_1(\boldsymbol{k})+E_2(\boldsymbol{k}) \right]_{\mathrm{min}}$,
where a peak shows up due to an enhancement of density of states.  
On the other hand, $\text{Re}(\sigma_H)$ is related to $\text{Im}(\sigma_H)$ through the Kramers-Kronig relation,
\begin{align}
& \text{Re} \left[ \sigma_H(\omega) \right] = \frac{1}{\pi} \mathcal{P} \int_{-\infty}^\infty \frac{\text{Im} \left[ \sigma_H \left( \omega' \right) \right]}{\omega'-\omega} d\omega',
\label{eq:Kramers}
\end{align}
which means that if a peak appears in the imaginary part at a certain frequency, the real part will inevitably undergo significant changes in the nearby frequency range, and it is highly likely to also form a peak.

\begin{figure}[ht]
    \centering
    \begin{subfigure}[b]{0.235\textwidth}
        \centering
        \includegraphics[width=\textwidth]{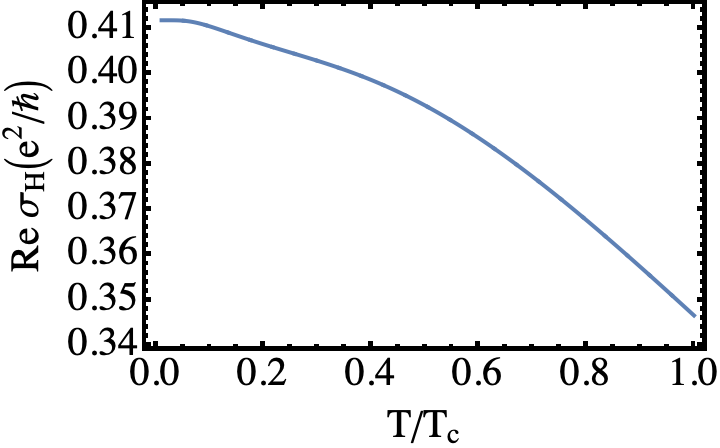}
        \caption{}
        \label{fig:sigmaHETRe}
    \end{subfigure}
    \hfill
    \begin{subfigure}[b]{0.235\textwidth}
        \centering
        \includegraphics[width=\textwidth]{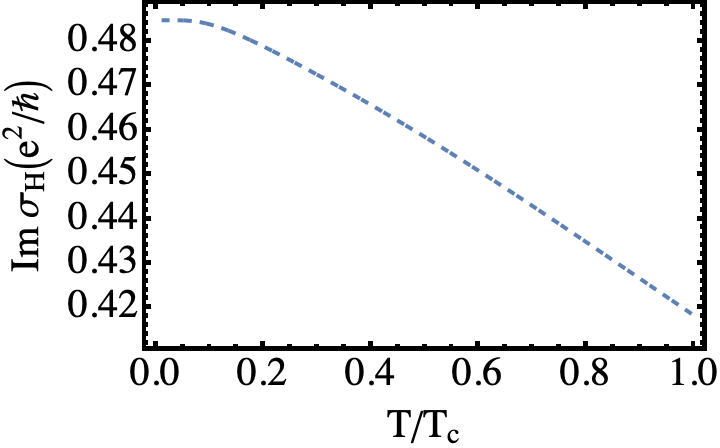}
        \caption{}
        \label{fig:sigmaHETIm}
    \end{subfigure}
   \captionsetup{justification=raggedright}
\caption{Real (panel (a)) and imaginary (panel (b)) parts of the anomalous AC Hall conductivity for $s+d_{x^2-y^2}e^{i\phi_1}+d_{xy}e^{i\phi_2}$ pairing  as functions of the temperature.
Same order parameters and chemical potential are taken as in Fig. \ref{fig:sigmaH}, and the frequency is set to be $\hbar \omega=0.9t$.}
\label{fig:sigmaHT}
\end{figure}

In addition, the temperature dependence of the anomalous AC Hall conductivity without any symmetry operations is shown in Fig. \ref{fig:sigmaHT}.
For clearer numerical variation, the frequency is chosen near the peak value of $\sigma_H$ in Fig. \ref{fig:sigmaH}(a),
i.e., $\hbar \omega=0.9t$.
As the temperature increases from $0$ to the superconducting critical temperature $T_c$, both the real and imaginary parts of $\sigma_H$ decrease gradually, with a slow decay near $T=0$ and a faster decay close to $T_c$.
We note that for the $\sigma_H$ curves with any symmetry operation, fixing the frequency at an arbitrary value should yield a temperature dependence of $\sigma_H$ similar to that shown in Fig. \ref{fig:sigmaHT}.

A non-vanishing $\sigma_H(\omega)$ leads to Kerr effect in the material, 
which can be used as an experimental signature for detecting time reversal symmetry breaking. 
When polarized light is incident on the surface of a chiral topological superconductor, the polarization direction of the reflected light undergoes rotation due to the nonzero AC Hall conductivity of the material.
For thick samples ($h \gg \lambda$), the Kerr angle $\theta_K$ depends on $\sigma_H(\omega)$ as follows \cite{Argyres1955}:
\begin{equation}
\theta_K(\omega)=\frac{2\pi}{d\omega}\text{Im}\left( \frac{\sigma_H(\omega)}{n\left(n^2-1\right)} \right),
\end{equation}
where $\lambda$ is the wavelength of the incident light, and $d$ denotes the separation of monolayer pairs.
And for thin samples ($h \ll \lambda$), the Kerr angle is given by \cite{Tse2011},
\begin{equation}
\theta_K(\omega)=\text{Re}\arctan{\left( \frac{-\sigma_H}{\sigma_{xx}+4\pi \left(\sigma_{xx}^2+\sigma_H^2 \right)} \right)},
\end{equation}
where $\sigma_{xx}$ is the longitudinal optical conductivity.

\begin{figure}[ht]
    \centering
    \begin{subfigure}[b]{0.24\textwidth}
        \centering
        \includegraphics[width=\textwidth]{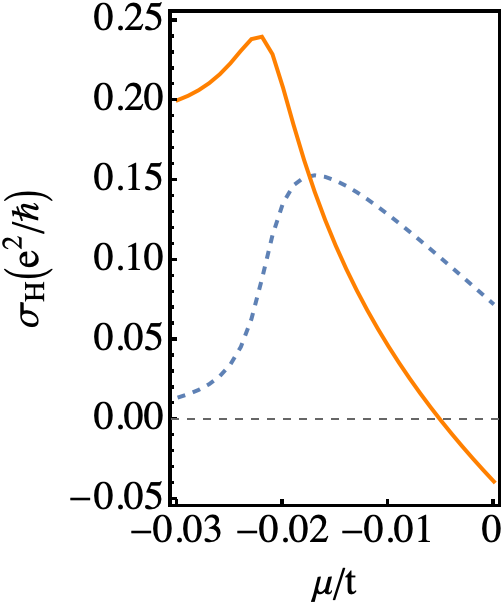}
        \caption{}
        \label{fig:sigmaHudds}
    \end{subfigure}
    \hfill
    \begin{subfigure}[b]{0.235\textwidth}
        \centering
        \includegraphics[width=\textwidth]{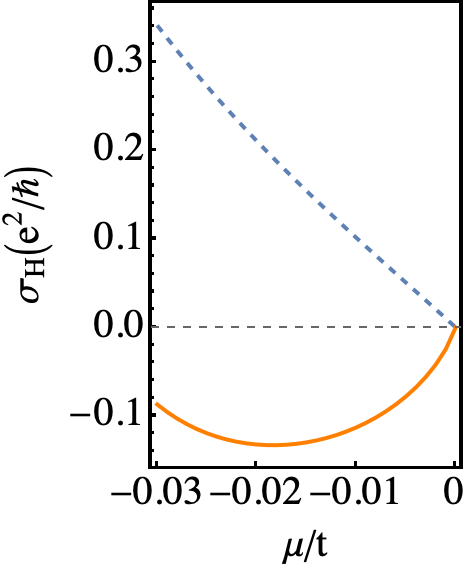}
        \caption{}
        \label{fig:sigmaHudd}
    \end{subfigure}
   \captionsetup{justification=raggedright}
\caption{Real (solid line) and imaginary (dashed line) parts of the anomalous AC Hall conductivity for (a) $s+d_{x^2-y^2}e^{i\phi_1}+d_{xy}e^{i\phi_2}$ pairing, and (b) $d_{x^2-y^2}+id_{xy}$ pairing gap function, as functions of chemical potential.
In panel (a), the same parameters for the tri-component pairing are used as in Fig. \ref{fig:Egap},
while in panel (b), $|\psi_{x^2-y^2}|=|\psi_{xy}|=0.1t.$
It is worth noting that in Panel (a), $\sigma_H$ for the tri-component pairing is nonzero at $\mu=0$, with a value of $(-0.0387+0.0723i)e^2/\hbar$,
while $\sigma_H$ for the $d+id$ pairing completely vanishes at $\mu=0$.
In all plots, we set $k_BT=0.01t$ with mixed $\hbar \omega=0.35t$.
}
\label{fig:sigmaHu}
\end{figure}

\subsection{Comparison with $d+id$ pairing and breaking of rotational symmetry}

We emphasize that the behavior of AC Hall conductivity for the tri-component $s + d_{x^2-y^2} e^{i\phi_1} + d_{xy} e^{i\phi_2}$ pairing exhibits notable differences compared with the chiral $d$-wave one $d_{x^2-y^2} + id_{xy}$.
In the chiral $d$-wave case, $\sigma_H$ vanishes at the Dirac point where the chemical potential $\mu = 0$.
On the other hand, $\sigma_H$ is non-vanishing even at the Dirac point for the $s + d_{x^2-y^2} e^{i\phi_1} + d_{xy} e^{i\phi_2}$ pairing, because of the presence of the $\Delta_s \left( \epsilon_x \text{Im} \Delta_y -\epsilon_y\text{Im} \Delta_x \right)$ term in the numerator of Eq. (\ref{eq:sigmaH}), as shown in Fig. \ref{fig:sigmaHu}.

It is noted that, for the parameter choices in the figure, the anomalous AC Hall conductivity of the tri-component pairing exhibits a single peak in both the real and imaginary parts within the range $-0.03t \leq \mu \leq 0$, whereas the curve for the $d+id$ pairing gradually approaches zero.
The origin of the peak in Fig. \ref{fig:sigmaHu}(a) is essentially the same as those in Fig. \ref{fig:sigmaH}.
Specifically, when $\hbar\omega = 0.35t$ is fixed, the peak position $\mu_{\mathrm{peak}}$ satisfies $(E_1+E_2)_{\mathrm{min}} = 0.35t=\hbar\omega$.
At this threshold, $\text{Im}(\sigma_H)$ exhibits a peak due to an enhancement of the density of states,
and $\text{Re}(\sigma_H)$ follows via the Kramers–Kronig relation.
For $|\mu|>|\mu_{\mathrm{peak}}|$, one has $(E_1+E_2)_{\mathrm{min}}>0.35t$, and $\sigma_H$ has not yet attained its maximum.
For $|\mu|<|\mu_{\mathrm{peak}}|$, $(E_1+E_2)_{\mathrm{min}}<0.35t$, leading to a gradual decrease after the peak, including a region where $\text{Re}(\sigma_H)<0$, consistent with Fig. \ref{fig:sigmaH}.
By contrast, for the $d+id$ pairing in Fig. \ref{fig:sigmaHu}(b), at $\mu=-0.03t$ one already has $(E_1+E_2)_{\mathrm{min}}<0.35t$, and this inequality persists as $|\mu|$ decreases further.
Consequently, $\sigma_H$ exhibits no peak and gradually approaches zero.

Furthermore, unlike the chiral $d$-wave pairing, the AC Hall conductivity for the  $s + d_{x^2-y^2} e^{i\phi_1} + d_{xy} e^{i\phi_2}$ pairing is not invariant under the $C_6$ rotational operation, since $C_6$ symmetry is spontaneously broken in the tri-component case. 
Fig. \ref{fig:sigmaH} (b) and (c) show the line-shapes of $\text{Re} \left[\sigma_H(\omega) \right]$ and $\text{Im} \left[\sigma_H(\omega) \right]$ when the coordinate axes are rotated by $\pi/3$ and $2\pi/3$ angles around the $z$-axis, respectively. 
Namely, the $\hat{x}$- and $\hat{y}$-directions in Eq. (\ref{eq:sigmaH}) for calculating $\sigma_H(\omega)$ are replaced by $\hat{x}^\prime$- and $\hat{y}^\prime$-directions, where $(\hat{x}^\prime, \hat{y}^\prime)$ are obtained from $(\hat{x}, \hat{y})$ by a rotation of angle $\pi/3$ for Fig. \ref{fig:sigmaH} (b) and $2\pi/3$ for Fig. \ref{fig:sigmaH} (c). 
It is evident from Fig. \ref{fig:sigmaH} (a,b,c) that the line-shapes are different for the three setups, indicating an anisotropy in the AC Hall response of the system along different directions. 
Such anisotropy can be used as an experimental probe to distinguish between chiral $d+id$ pairing and the tri-component $s + d_{x^2-y^2} e^{i\phi_1} + d_{xy} e^{i\phi_2}$ pairing.


\section{Fractional vortices in magnetic fields}
\label{sec:fractional_vortex}

In this section, we discuss another physical property of the tri-component pairing with mixed $d_{x^2-y^2}$-, $d_{xy}$-, and $s$-wave symmetries -- fractional vortices, which occur in multi-component pairing systems \cite{Babaev2002}. 

To study vortex structures in superconductors, the spatial gradient terms of order parameters and magnetic fields  need to be included in the Ginzburg-Landau free energy.
By incorporating these effects, the free energy  functional can be written as:
\begin{align}
F = &F_s^{(0)}+F_d^{(0)}+F^{(4)}+\frac{\mathbf{B}^2}{2}+\frac{1}{2m_s}\left|\left(\nabla+2ie\mathbf{A} \right)\psi_s \right|^2 \notag\\
&+ \frac{1}{2m_1}\left|\left(\nabla+2ie\mathbf{A} \right)\psi_1 \right|^2 + \frac{1}{2m_2}\left|\left(\nabla+2ie\mathbf{A} \right)\psi_2 \right|^2,
\label{eq:F_A}
\end{align}
in which the expressions for $F_s^{(0)}$ and $F_d^{(0)}$ are the same as in Eq. \eqref{eq:F0}, and $F^{(4)}$ is given in Eq. \eqref{eq:F4}.
Here, $m_{s,1,2}$ are the effective masses of the Cooper pairs associated with the order parameters $\psi_{s,1,2}$,
$e$ is the electron charge, and the order parameters are coupled to the vector potential $\mathbf{A}$ via minimal coupling.
Performing the functional derivative of Eq. (\ref{eq:F_A}) with respect to the vector potential $\mathbf{A}$ (for details, see Appendix \ref{app:supercurrent}), we obtain the expression of the  supercurrent as
\begin{align}
\mathbf{J}=-4e\rho^2 & \left[ \cos^2{\left(\frac{\theta}{2}\right)} \nabla \phi_s+\sin^2{\left(\frac{\theta}{2}\right)} \cos^2{\left(\frac{\gamma}{2}\right)} \nabla \phi_1 \right. \notag\\
& \left. + \sin^2{\left(\frac{\theta}{2}\right)} \sin^2{\left(\frac{\gamma}{2}\right)} \nabla \phi_2+2e\mathbf{A} \right],
\label{eq:J}
\end{align}
in which 
\begin{eqnarray}
\rho^2=\frac{\left|\psi_s \right|^2}{2m_s} + \frac{\left|\psi_1 \right|^2}{2m_1} + \frac{\left|\psi_2\right|^2}{2m_2}, 
\end{eqnarray}
and $\theta$ and $\gamma$ are given by:
\begin{align}
& \left|\psi_s\right|=\sqrt{2m_s}\rho \cos{\left(\frac{\theta}{2}\right)}, \notag\\
& \left|\psi_1\right|=\sqrt{2m_1}\rho \sin{\left(\frac{\theta}{2}\right)} \cos{\left(\frac{\gamma}{2}\right)}, \notag\\
& \left|\psi_2\right|=\sqrt{2m_2}\rho \sin{\left(\frac{\theta}{2}\right)} \sin{\left(\frac{\gamma}{2}\right)}.
\label{eq:psi_rho}
\end{align}

In the far-field region, i.e., at distances away from the vortex core much larger than the magnetic penetration length $\lambda$,
the supercurrent vanishes. Therefore, integrating over a closed path $\sigma$ around the vortex core in the far-field region gives:
\begin{equation}
\oint_{\sigma} d\boldsymbol{l} \cdot \mathbf{J}=0.
\label{eq:J_0}
\end{equation}
Then we arrive at the following equation for the magnetic flux $\Phi=\oint_\sigma d\boldsymbol{l} \cdot \mathbf{A}$ carried by the vortex:
\begin{align}
\Phi=-\frac{1}{2e} & \left[\cos^2{\left(\frac{\theta}{2}\right)} \Delta \phi_s+\sin^2{\left(\frac{\theta}{2}\right)} \cos^2{\left(\frac{\gamma}{2}\right)} \Delta \phi_1 \right. \notag\\
& \left. + \sin^2{\left(\frac{\theta}{2}\right)} \sin^2{\left(\frac{\gamma}{2}\right)} \Delta \phi_2 \right],
\label{eq:magnetic_flux}
\end{align}
where $\Delta \phi_{s,1,2}=\oint_\sigma d\boldsymbol{l} \cdot \nabla \phi_{s,1,2}$ are the phase windings of the order parameters.
It then follows that depending on the values of $\Delta\phi_s$, $\Delta\phi_1$, and $\Delta\phi_2$, vortices in tri-component pairing superconducting system can carry either integer or arbitrary fractional magnetic flux quanta.
Specifically, when $\Delta\phi_s = \Delta\phi_1 = \Delta\phi_2 = 2n\pi$, the magnetic flux of the vortex is $\Phi = -n\Phi_0$,
where $\Phi_0 = \pi/e$ denotes the standard flux quantum.
In this case, the vortices are the Abrikosov vortices of an ordinary superconductor, each carrying $n$ flux quanta.

If $\Delta\phi_1 = \Delta\phi_2=2n\pi$ and $\Delta\phi_s + \Delta\phi_1 = 0$, the magnetic flux becomes
\begin{eqnarray}
\Phi &= &-\frac{1}{2e} \left[-\cos^2{\left(\frac{\theta}{2}\right)}\Delta \phi_1+\sin^2{\left(\frac{\theta}{2}\right)}\Delta \phi_1\right]\nn\\
&=&\cos{\theta}n\Phi_0.
\label{eq:Phi_cos}
\end{eqnarray}
Since $\cos{\theta}$ can take arbitrary value, such a vortex can carry an arbitrary fraction of the magnetic flux quantum, similar to the case in two-component superconductors.
Likewise, if $\Delta\phi_1 = \Delta\phi_s=2n\pi$ and $\Delta\phi_1 + \Delta\phi_2 = 0$, one obtains
\begin{equation}
\Phi = -\left[\cos^2{\left(\frac{\theta}{2}\right)} + \sin^2{\left(\frac{\theta}{2}\right)} \cos{\gamma} \right]n\Phi_0,
\label{eq:fracvortex1}
\end{equation}
and if instead $\Delta\phi_2 = \Delta\phi_s=2n\pi$ and $\Delta\phi_1 + \Delta\phi_2 = 0$, the flux is
\begin{equation}
\Phi = -\left[\cos^2{\left(\frac{\theta}{2}\right)} - \sin^2{\left(\frac{\theta}{2}\right)} \cos{\gamma} \right]n\Phi_0.
\label{eq:fracvortex2}
\end{equation}
In either case, vortices can carry arbitrary fractional magnetic flux quanta.

More generally, for $\Delta\phi_s = 2k_s\pi$, $\Delta\phi_1 = 2k_1\pi$, and $\Delta\phi_2 = 2k_2\pi$, the flux carried by the vortex is given by:
\begin{eqnarray}
\left|\Phi \right|&=&\lambda_{k_sk_1k_2}\Phi_0,
\label{eq:vortex_ks12}
\end{eqnarray}
in which
\begin{eqnarray}
\lambda_{k_sk_1k_2}&=&k_s\cos^2{\left(\frac{\theta}{2}\right)} +k_1\sin^2{\left(\frac{\theta}{2}\right)} \cos^2{\left(\frac{\gamma}{2}\right)}\notag\\
 &&+ k_2\sin^2{\left(\frac{\theta}{2}\right)} \sin^2{\left(\frac{\gamma}{2}\right)},
\end{eqnarray}
or alternatively
\begin{eqnarray}
\lambda_{k_sk_1k_2}&=&\left( k_s\frac{\left|\psi_s \right|^2}{m_s}+k_1\frac{\left|\psi_1 \right|^2}{m_1}+k_2\frac{\left|\psi_2 \right|^2}{m_2} \right)\notag\\
&&\cdot\left( \frac{\left|\psi_s \right|^2}{m_s}+\frac{\left|\psi_1 \right|^2}{m_1}+\frac{\left|\psi_2 \right|^2}{m_2} \right)^{-1},
\end{eqnarray}
which is again a fractional vortex with arbitrary value of magnetic flux.


\section{conclusions}
\label{sec:conclusion}

In conclusion, we have investigated the existence of chiral topological superconductivity on a two-dimensional honeycomb lattice with tri-component pairing gap function of mixed $s$-, $d_{x^-y^2}$- and $d_{xy}$-wave symmetries.
Using a Ginzburg-Landau free energy analysis, the overall pairing gap function can be determined as $s+ d_{x^2-y^2}e^{i\phi_1} + d_{xy}e^{i\phi_2}$,
which spontaneously breaks the time reversal, rotational and reflectional symmetries. 
The symmetry-breaking pattern of the pairing configuration is $C_{6v} \times \mathbb{Z}_2^T \to C_2$,
leading to 12 degenerate solutions of the ground state pairing configuration.
Based on a microscopic model for the tri-component pairing on the honeycomb lattice,
the system is shown to be a fully gapped topological superconductor with nonzero Chern number and mid-gap edge states. 
Furthermore, the anomalous AC Hall conductivity is calculated to be non-vanishing, 
which breaks the $C_6$ rotational symmetry,
reflecting the anisotropic nature of the tri-component pairing gap function.
Fractional vortices are also discussed, which arises from the multi-component pairing structure of the system.

\section*{Acknowledgments}

Y.L., J.J., and W.Y. are supported by the startup funding at Nankai University. 
X.Z. acknowledgments the startup funding of Huazhong University of Science and Technology.
C.W. is supported by the National Natural Science Foundation of China under the Grants No. 12234016 and No.12174317.
This work has been supported by the New Cornerstone Science Foundation.

\appendix

\begin{widetext}

\section{G-L free energy analysis of a tri-component pairing function with $C_{6v}$ symmetry}
\label{app:glfree}

In this section, we give a quick review of the $C_{6v}$ group and G-L free energy.
Here $(x^2-y^2,xy)$ is an $E_2$-representation of the $C_{6v}$ group, and $A_1$-representation is symmetric under all operations.
The product rules for the $A_1$-, $A_2$- and $E_2$-representations of $C_{6v}$ and the corresponding example functions can be worked out as Table \ref{tab:C6v}:
\begin{table}[h!]
\centering
\begin{tabular}{|c|c|c|c|c|}
    \hline
    $C_{6v}$ & $A_1$ & $A_2$ & $E_2$ & functions \\
    \hline
    $A_1$ & $A_1$ & $A_2$ & $E_2$ & $z, x^2+y^2, z^2$ \\
    \hline
    $A_2$ & $~$ & $A_1$ & $E_2$ & $J_z$ \\
    \hline
    $E_2$ & $~$ & $~$ & $A_1+A_2+E_2$ & $(x^2-y^2, xy)$ \\
    \hline
\end{tabular}
\caption{Multiplication Table of irreducible representations of the $C_{6v}$ group}
\label{tab:C6v}
\end{table}

The two generators of the $C_{6v}$ group are $r = C_6$, and $f = l_1$, as defined in the main text.
The generator relation representation for $C_{6v}$ is $C_{6v} = \langle r,f | r^6 = f^2 = (rf)^6 = e \rangle$.
Write $d_1 = \hat{d}_{x^2-y^2}$, and $d_2 = \hat{d}_{xy}$.
The result of the action of the two group generators on this set of coordinates is
\begin{align}
C_6 \begin{pmatrix}
d_1 \\
d_2
\end{pmatrix}
&=
\begin{pmatrix}
-1/2 & \sqrt{3}/2  \\
-\sqrt{3}/2 & -1/2
\end{pmatrix}
\begin{pmatrix}
d_1 \\
d_2
\end{pmatrix},
\notag\\
l_1 \begin{pmatrix}
d_1 \\
d_2
\end{pmatrix}
&=
\begin{pmatrix}
1 & 0  \\
0 & -1
\end{pmatrix}
\begin{pmatrix}
d_1 \\
d_2
\end{pmatrix}.
\label{eq:generator}
\end{align}

Since the coordinate of the linear term of $d$ is $(d_1, d_2)$, when we come to the quadratic term of $d$, $A_1$ is represented as $|d_1|^2+|d_2|^2$, $A_2$ is represented as $i(d_1^*d_2-d_2^*d_1)$, and $E$ is represented as $(|d_1|^2-|d_2|^2, d_1^*d_2+d_2^*d_1)$.

\subsection{Quadratic terms in free energy of $d$-wave}

$(d_1,d_2)$ forms an $E_2$-representation of $C_{6v}$ group, and $E_2 \times E_2 = A_1+A_2+E_2$,
there is only one $C_{6v}$-invariant combination, $A_1$.
Thus we have only one term in the free energy of $d$-wave up to quadratic level, i.e.,
\begin{equation}
 f^{(2)} = |\psi_1|^2 + |\psi_2|^2.
\end{equation}


\subsection{Cubic terms in free energy of $d$-wave}

Although the free energy only contains even-order terms of the order parameters,
the presence of an isotropic $s$-wave order parameter, in addition to the $d$-wave order parameters,
allows the $d$-wave component to take cubic terms.
These cubic terms, together with the first order terms of the $s$-wave order parameter,
contribute to the quartic terms in the free energy.
\begin{equation}
E_2 \times E_2 \times E_2 = \left( A_1+A_2+E_2 \right) \times E_2,
\label{eq:cubic}
\end{equation}
where the first $E_2$ on the right-hand side of Eq. \eqref{eq:cubic} is the quadratic term of $d$-wave components,
and the second $E_2$ is the linear term.

The action of $C_6$ and $l_1$ on the $E_2$-representation which describes the linear term of the $d$-wave components is given by Eq. \eqref{eq:generator},
while the transformations of the $E_2$-representation which describes the quadratic term of the $d$-wave components are given by
\begin{align}
C_6 \begin{pmatrix}
|d_1|^2-|d_2|^2 \\
d_1^*d_2+d_2^*d_1
\end{pmatrix}
&=
\begin{pmatrix}
-1/2 & -\sqrt{3}/2  \\
\sqrt{3}/2 & -1/2
\end{pmatrix}
\begin{pmatrix}
|d_1|^2-|d_2|^2 \\
d_1^*d_2+d_2^*d_1
\end{pmatrix},
\notag\\
l_1 \begin{pmatrix}
|d_1|^2-|d_2|^2 \\
d_1^*d_2+d_2^*d_1
\end{pmatrix}
&=
\begin{pmatrix}
1 & 0  \\
0 & -1
\end{pmatrix}
\begin{pmatrix}
|d_1|^2-|d_2|^2 \\
d_1^*d_2+d_2^*d_1
\end{pmatrix}.
\label{eq:d1d2}
\end{align}
It is evident that $(d_1,d_2)\sigma_z$ transform in the same way as $(|d_1|^2-|d_2|^2,d_1^*d_2+d_2^*d_1)$ under the $C_{6v}$ group, where $\sigma_z$ is the Pauli matrix.
Thus, there is one cubic term of $d$-wave order parameters,
\begin{equation}
f^{(3)} = \psi_1 \left( |\psi_1|^2-|\psi_2|^2 \right) - \psi_2 \left( \psi_1^*\psi_2+\psi_2^*\psi_1 \right).
\end{equation}


\subsection{Quartic terms in free energy of $d$-wave}

Up to quartic terms, we need to consider the product $(E_2 \times E_2) \times (E_2 \times E_2)$,
\begin{align}
(E_2 \times E_2) \times (E_2 \times E_2) &= ( A_1+A_2+E_2) \times ( A_1+A_2+E_2) \notag\\
&= A_1+A_2+E_2+A_2+A_1+E_2 + E_2+E_2+(A_1+A_2+E_2).
\end{align}
So there should be three extra $C_{6v}$-invariant terms in the free energy up to quartic terms:
\begin{align}
& A_1 \times A_1: f_1^{(4)} = (|\psi_1|^2 + |\psi_2|^2)^2, \notag\\
& A_2 \times A_2: f_2^{(4)} = -(\psi_1^*\psi_2 - \psi_2^*\psi_1)^2, \notag\\
& E_2 \times E_2: f_3^{(4)} = (|\psi_1|^2 - |\psi_2|^2)^2 + (\psi_1^*\psi_2 + \psi_2^*\psi_1)^2.
\end{align}
However, 
$f_3^{(4)} = |\psi_1|^4 + |\psi_2|^4 + \psi_1^{*2}\psi_2^2 + \psi_2^{*2}\psi_1^2= f_1^{(4)} - f_2^{(4)}$,
thus, there are only two extra linearly independent terms up to quartic level.


If we include the mixture with an $s$-wave pairing order parameter, the overall free energy up to the quartic order is
\begin{eqnarray}
F&=&\alpha_d \left( |\psi_1|^2 + |\psi_2|^2 \right) + \beta_d \left( |\psi_1|^2 + |\psi_2|^2 \right)^2 +  \alpha_s |\psi_s|^2 + \beta_s |\psi_s|^4 \notag\\
&+&\gamma |\psi_s|^2 \left( |\psi_1|^2 + |\psi_2|^2 \right) + g_{dd} \left( \psi_1^*\psi_2 - \psi_1 \psi_2^* \right)^2
+ g_{sd} \left[ \psi_s^2 \left( \psi_1^{*2}+\psi_2^{*2} \right) + \psi_s^{*2} \left( \psi_1^2+\psi_2^2 \right) \right] \notag\\
&+&g'_{sd} \left[ \left( \psi_s^* \psi_1 + \psi_s \psi_1^* \right) \left(|\psi_1|^2-|\psi_2|^2 \right) - \left( \psi_s^* \psi_2 + \psi_s \psi_2^* \right) \left( \psi_1^*\psi_2+\psi_1\psi_2^* \right) \right].
\label{eq:Fdds}
\end{eqnarray}

\section{$d$-wave symmetry of the pairing functions $\Delta_{x^2-y^2}$ and $\Delta_{xy}$}
\label{app:dwave}

For simplicity, the pairing term involving $\Delta_{x^2-y^2}$ and $\Delta_{xy}$ separately can be considered as:
\begin{equation}
\Delta_d(\boldsymbol{k})=\Delta_{x^2-y^2}(\boldsymbol{k})e^{i\phi_1}+\Delta_{xy}(\boldsymbol{k})e^{i\phi_2},
\label{eq:Deltad}
\end{equation}
where the definitions of $\Delta_{x^2-y^2}$ and $\Delta_{xy}$ are given in Eq. \eqref{eq:Deltas12} in the main text.
Let
\begin{eqnarray}
 g_1&=&|\psi_1|e^{i\phi_1}\left[\cos{\left(k_xa\right)}-\cos{\left(\frac{1}{2}k_xa\right)}\cos{\left(\frac{\sqrt{3}}{2}k_ya\right)}\right], \notag\\
g_2&=&|\psi_2|e^{i\phi_2}\left[-\sqrt{3}\sin{\left(\frac{1}{2}k_xa\right)}\sin{\left(\frac{\sqrt{3}}{2}k_ya\right)}\right], \notag\\
-h_1&=&-|\psi_1|e^{i\phi_1}\left[\sin{\left(k_xa\right)}+\sin{\left(\frac{1}{2}k_xa\right)}\cos{\left(\frac{\sqrt{3}}{2}k_ya\right)}\right], \notag\\
h_2&=&|\psi_2|e^{i\phi_2}\left[\sqrt{3}\cos{\left(\frac{1}{2}k_xa\right)}\sin{\left(\frac{\sqrt{3}}{2}k_ya\right)}\right].
\end{eqnarray}
In this case, we have
\begin{eqnarray}
\Delta_{x^2-y^2}(\boldsymbol{k})e^{i\phi_1}&=&g_1\sigma_x+(-h_1)\sigma_y,\notag\\
\Delta_{xy}(\boldsymbol{k})e^{i\phi_2}&=&g_2\sigma_x+h_2\sigma_y.
\end{eqnarray}
The pairing operator that only involves $\Delta_{x^2-y^2}$ and $\Delta_{xy}$ can be written as
\begin{align}
\hat{\Delta}_d&=\sum_{\boldsymbol{k}} \Psi_{\boldsymbol{k}}^{\dagger}
\begin{pmatrix}
0 & \Delta_d(\boldsymbol{k}) \\
 \Delta_d^\dagger(\boldsymbol{k}) & 0
\end{pmatrix}
\Psi_{\boldsymbol{k}},
\end{align}
or more explicitly,
\begin{align}
\hat{\Delta}_d&= \sum_{\boldsymbol{k}}
(c_{\boldsymbol{k}_A}^\dagger~c_{\boldsymbol{k}_B}^\dagger~c_{-\boldsymbol{k}_A}~c_{-\boldsymbol{k}_B})
\begin{pmatrix}
0 & \left(g_1+g_2\right)\sigma_x+\left(-h_1+h_2\right)\sigma_y \\
\left[ \left(g_1+g_2\right)\sigma_x+\left(-h_1+h_2\right)\sigma_y \right]^\dagger & 0
\end{pmatrix}
\begin{pmatrix}
c_{\boldsymbol{k}_A} \\
c_{\boldsymbol{k}_B} \\
c_{-\boldsymbol{k}_A}^\dagger \\
c_{-\boldsymbol{k}_B}^\dagger
\end{pmatrix},
\end{align}
where the spin indices are omitted because we focus on the momentum, with upward spin always paired with positive momentum and downward spin with negative momentum.
Consider the two parts that include $\sigma_x$ and $\sigma_y$ separately,
\begin{equation}
\Delta_d(\boldsymbol{k})=\Delta_{dx}(\boldsymbol{k})+\Delta_{dy}(\boldsymbol{k}),
\end{equation}
in which $\Delta_{dx}(\boldsymbol{k})=\left(g_1+g_2\right)\sigma_x$, and $\Delta_{dy}(\boldsymbol{k})=\left(-h_1+h_2\right)\sigma_y$.

The action of $C_6$ rotation on $\hat{\Delta}_{dx}$ is
\begin{align}
\hat{U}(C_6) \hat{\Delta}_{dx} \hat{U}^\dagger(C_6)
&= \sum_{\boldsymbol{k}}
\hat{U}(C_6) \Psi_{\boldsymbol{k}}^\dagger \hat{U}^\dagger(C_6) \hat{U}(C_6)
\begin{pmatrix}
0 & \left(g_1+g_2\right)\sigma_x \\
\left[ \left(g_1+g_2\right)\sigma_x \right]^\dagger & 0
\end{pmatrix}
\hat{U}^\dagger(C_6)
\hat{U}(C_6) \Psi_{\boldsymbol{k}} \hat{U}^\dagger(C_6) \notag\\
&= \sum_{\boldsymbol{k}}
(c_{\boldsymbol{k}_A}^\dagger~c_{\boldsymbol{k}_B}^\dagger~c_{-\boldsymbol{k}_A}~c_{-\boldsymbol{k}_B})
\begin{pmatrix}
0 & \left(g'_1+g'_2\right)\sigma_x \\
\left[ \left(g'_1+g'_2\right)\sigma_x \right]^\dagger & 0
\end{pmatrix}
\begin{pmatrix}
c_{\boldsymbol{k}_A} \\
c_{\boldsymbol{k}_B} \\
c_{-\boldsymbol{k}_A}^\dagger \\
c_{-\boldsymbol{k}_B}^\dagger
\end{pmatrix},
\end{align}
in which
\begin{align}
g'_1&=|\psi_1|e^{i\phi_1}\left \{-\frac{1}{2} \left[\cos{\left(k_xa\right)}-\cos{\left(\frac{1}{2}k_xa\right)}\cos{\left(\frac{\sqrt{3}}{2}k_ya\right)}\right] - \frac{\sqrt{3}}{2} \left[-\sqrt{3}\sin{\left(\frac{1}{2}k_xa\right)}\sin{\left(\frac{\sqrt{3}}{2}k_ya\right)}\right] \right \},
\notag\\
g'_2&=|\psi_2|e^{i\phi_2}\left \{\frac{\sqrt{3}}{2} \left[\cos{\left(k_xa\right)}-\cos{\left(\frac{1}{2}k_xa\right)}\cos{\left(\frac{\sqrt{3}}{2}k_ya\right)}\right] - \frac{1}{2} \left[-\sqrt{3}\sin{\left(\frac{1}{2}k_xa\right)}\sin{\left(\frac{\sqrt{3}}{2}k_ya\right)}\right] \right \}.
\end{align}

The action of the $l_1$ mirror reflection on $\hat{\Delta}_{dx}$ is
\begin{align}
\hat{U}(l_1) \hat{\Delta}_{dx} \hat{U}^\dagger(l_1)
&= \sum_{\boldsymbol{k}}
\hat{U}(l_1) \Psi_{\boldsymbol{k}}^\dagger \hat{U}^\dagger(l_1) \hat{U}(l_1)
\begin{pmatrix}
0 & \left(g_1+g_2\right)\sigma_x \\
\left[ \left(g_1+g_2\right)\sigma_x \right]^\dagger & 0
\end{pmatrix}
\hat{U}^\dagger(l_1)
\hat{U}(l_1) \Psi_{\boldsymbol{k}} \hat{U}^\dagger(l_1) \notag\\
&= \sum_{\boldsymbol{k}}
(c_{\boldsymbol{k}_A}^\dagger~c_{\boldsymbol{k}_B}^\dagger~c_{-\boldsymbol{k}_A}~c_{-\boldsymbol{k}_B})
\begin{pmatrix}
0 & \left(g''_1+g''_2\right)\sigma_x \\
\left[ \left(g''_1+g''_2\right)\sigma_x \right]^\dagger & 0
\end{pmatrix}
\begin{pmatrix}
c_{\boldsymbol{k}_A} \\
c_{\boldsymbol{k}_B} \\
c_{-\boldsymbol{k}_A}^\dagger \\
c_{-\boldsymbol{k}_B}^\dagger
\end{pmatrix},
\end{align}
where
\begin{eqnarray}
g''_1 &=& |\psi_1|e^{i\phi_1}\left[ \cos{\left(k_xa\right)}-\cos{\left(\frac{1}{2}k_xa\right)}\cos{\left(-\frac{\sqrt{3}}{2}k_ya\right)}\right] = g_1, \notag\\
g''_2 &=& |\psi_2|e^{i\phi_2}\left[ -\sqrt{3} \sin{\left(\frac{1}{2}k_xa \right)} \sin{\left( -\frac{\sqrt{3}}{2}k_ya \right)} \right]=-g_2.
\end{eqnarray}
Thus, $\left( g_1\sigma_x, g_2\sigma_x \right)$ forms an $E_2$-representation of the $C_{6v}$ group.

On the other hand, the action of $C_6$ rotation on $\hat{\Delta}_{dy}$ is
\begin{align}
\hat{U}(C_6) \hat{\Delta}_{dy} \hat{U}^\dagger(C_6)
&= \sum_{\boldsymbol{k}}
\hat{U}(C_6) \Psi_{\boldsymbol{k}}^\dagger \hat{U}^\dagger(C_6) \hat{U}(C_6)
\begin{pmatrix}
0 & \left(-h_1+h_2\right)\sigma_y \\
\left[ \left(-h_1+h_2\right)\sigma_y \right]^\dagger & 0
\end{pmatrix}
\hat{U}^\dagger(C_6)
\hat{U}(C_6) \Psi_{\boldsymbol{k}} \hat{U}^\dagger(C_6) \notag\\
&= \sum_{\boldsymbol{k}}
(c_{\boldsymbol{k}_A}^\dagger~c_{\boldsymbol{k}_B}^\dagger~c_{-\boldsymbol{k}_A}~c_{-\boldsymbol{k}_B})
\begin{pmatrix}
0 & \left(-h'_1+h'_2\right)\sigma_y \\
\left[ \left(-h'_1+h'_2\right)\sigma_y \right]^\dagger & 0
\end{pmatrix}
\begin{pmatrix}
c_{\boldsymbol{k}_A} \\
c_{\boldsymbol{k}_B} \\
c_{-\boldsymbol{k}_A}^\dagger \\
c_{-\boldsymbol{k}_B}^\dagger
\end{pmatrix},
\end{align}
in which
\begin{align}
-h'_1&=-|\psi_1|e^{i\phi_1} \left \{-\frac{1}{2}\left[\sin(k_xa)+\sin{\left(\frac{1}{2}k_xa\right)}\cos{\left(\frac{\sqrt{3}}{2}k_ya\right)}\right] + \frac{\sqrt{3}}{2}\left[\sqrt{3}\cos{\left(\frac{1}{2}k_xa\right)} \sin{\left(\frac{\sqrt{3}}{2}k_ya \right)} \right] \right \}, \notag\\
h'_2 &=|\psi_2|e^{i\phi_2} \left \{-\frac{\sqrt{3}}{2}\left[\sin(k_xa)+\sin{\left(\frac{1}{2}k_xa\right)}\cos{\left(\frac{\sqrt{3}}{2}k_ya\right)}\right] - \frac{1}{2}\left[\sqrt{3}\cos{\left(\frac{1}{2}k_xa\right)} \sin{\left(\frac{\sqrt{3}}{2}k_ya \right)} \right] \right \}.
\end{align}

The result of the $l_1$ mirror reflection acting on $\hat{\Delta}_{dy}$ is
\begin{align}
\hat{U}(l_1) \hat{\Delta}_{dy} \hat{U}^\dagger(l_1)
&= \sum_{\boldsymbol{k}}
\hat{U}(l_1) \Psi_{\boldsymbol{k}}^\dagger \hat{U}^\dagger(l_1) \hat{U}(l_1)
\begin{pmatrix}
0 & \left(-h_1+h_2\right)\sigma_y \\
\left[ \left(-h_1+h_2\right)\sigma_y \right]^\dagger & 0
\end{pmatrix}
\hat{U}^\dagger(l_1)
\hat{U}(l_1) \Psi_{\boldsymbol{k}} \hat{U}^\dagger(l_1) \notag\\
&= \sum_{\boldsymbol{k}}
(c_{\boldsymbol{k}_A}^\dagger~c_{\boldsymbol{k}_B}^\dagger~c_{-\boldsymbol{k}_A}~c_{-\boldsymbol{k}_B})
\begin{pmatrix}
0 & \left(-h''_1+h''_2\right)\sigma_y \\
\left[ \left(-h''_1+h''_2\right)\sigma_y \right]^\dagger & 0
\end{pmatrix}
\begin{pmatrix}
c_{\boldsymbol{k}_A} \\
c_{\boldsymbol{k}_B} \\
c_{-\boldsymbol{k}_A}^\dagger \\
c_{-\boldsymbol{k}_B}^\dagger
\end{pmatrix},
\end{align}
where
\begin{eqnarray}
-h''_1&=&-|\psi_1|e^{i\phi_1}\left[ \sin{\left(k''_xa\right)}+\sin{\left(\frac{1}{2}k''_xa\right)}\cos{\left(-\frac{\sqrt{3}}{2}k''_ya\right)}\right] = -h_1, \notag\\
h''_2&=&|\psi_2|e^{i\phi_2}\left[ \sqrt{3} \cos{\left(\frac{1}{2}k''_xa \right)} \sin{\left( -\frac{\sqrt{3}}{2}k''_ya \right)} \right] = -h_2.
\end{eqnarray}
Therefore, $\left( -h_1\sigma_y, h_2\sigma_y \right)$ also forms an $E_2$-representation of the $C_{6v}$ group.

Since $(\Delta_{x^2-y^2}, \Delta_{xy})$ is a linear combination of $(g_1\sigma_x, g_2\sigma_x)$ and $(-h_1\sigma_y, h_2\sigma_y)$,
$(\Delta_{x^2-y^2}, \Delta_{xy})$ is also an $E_2$-representation of the $C_{6v}$ group and thus transforms in the same way as $(d_{x^2-y^2}, d_{xy})$.

\section{Anomalous AC Hall conductivity under the mirror reflection symmetries of $C_{6v}$}
\label{app:mirror}

From Eq. \eqref{eq:pixy} in the main text, the current-current correlator is
\begin{equation}
\pi_{xy}(\boldsymbol{q},\omega)=\int_0^\infty dte^{i\omega t}\left \langle \left[ \hat{J}_x^\dagger(\boldsymbol{q},t), \hat{J}_y(\boldsymbol{q},0) \right] \right \rangle,
\label{eq:pixy1}
\end{equation}
where $\hat{J}_\alpha=e\sum_{\boldsymbol{k}} \Psi_{\boldsymbol{k}}^\dagger \hat{v}_\alpha \Psi_{\boldsymbol{k}}$,
and $\hat{v}_\alpha=(\sigma_z \otimes \sigma_0) (1/\hbar) \partial_{k_\alpha}h_{\boldsymbol{k}}^0$, $\alpha=x,y$, i.e,
\begin{eqnarray}
\hat{v}_\alpha& =&
\begin{pmatrix}
1 & 0 & 0 & 0 \\
0 & 1 & 0 & 0 \\
0 & 0 & -1 & 0 \\
0 & 0 & 0 & -1 
\end{pmatrix}
\frac{1}{\hbar} ~ \partial_{k_\alpha}
 \begin{pmatrix}
-\mu & \epsilon_x-i\epsilon_y & 0 & 0 \\
\epsilon_x+i\epsilon_y & -\mu & 0 & 0 \\
0 & 0 & \mu & -\epsilon_x+i\epsilon_y \\
0 & 0 & -\epsilon_x-i\epsilon_y & \mu
\end{pmatrix} \notag\\
&=&
 \begin{pmatrix}
0 & v_\alpha & 0 & 0 \\
v_\alpha^* & 0 & 0 & 0 \\
0 & 0 & 0 & v_\alpha \\
0 & 0 & v_\alpha^* & 0
\end{pmatrix},
\end{eqnarray}
in which $v_\alpha=(1/\hbar) \partial_{k_\alpha} \left( \epsilon_x-i\epsilon_y \right)$, $v_\alpha^*=(1/\hbar) \partial_{k_\alpha} \left( \epsilon_x+i\epsilon_y \right)$.
For $v_x$ and $v_y$,
\begin{align}
& \partial_{k_x}\epsilon_x = ta\left[\sin \left( k_xa \right)+\frac{1}{2}\sin \left( \frac{1}{2}k_xa-\frac{\sqrt{3}}{2}k_ya \right)+\frac{1}{2}\sin \left( \frac{1}{2}k_xa+\frac{\sqrt{3}}{2}k_ya \right) \right], \notag\\
& \partial_{k_x}\epsilon_y = ta\left[\cos \left( k_xa \right)-\frac{1}{2}\cos \left( \frac{1}{2}k_xa-\frac{\sqrt{3}}{2}k_ya \right)-\frac{1}{2}\cos \left( \frac{1}{2}k_xa+\frac{\sqrt{3}}{2}k_ya \right) \right], \notag\\
& \partial_{k_y}\epsilon_x = -\frac{\sqrt{3}}{2}ta\left[\sin \left( \frac{1}{2}k_xa-\frac{\sqrt{3}}{2}k_ya \right) - \sin \left( \frac{1}{2}k_xa+\frac{\sqrt{3}}{2}k_ya \right) \right], \notag\\
& \partial_{k_y}\epsilon_y = \frac{\sqrt{3}}{2}ta\left[\cos \left( \frac{1}{2}k_xa-\frac{\sqrt{3}}{2}k_ya \right) - \cos \left( \frac{1}{2}k_xa+\frac{\sqrt{3}}{2}k_ya \right) \right].
\end{align}

Under mirror reflection transformation along $x$-axis, i.e, the $l_1$-reflection defined in Fig. \ref{fig:C6vgroup} in the main text,
the above partial derivatives transform as
\begin{eqnarray}
 l_1 \left( \partial_{k_x}\epsilon_x \right) &=& \partial_{k_x}\epsilon_x, \notag\\
  l_1 \left( \partial_{k_x}\epsilon_y \right) &=& \partial_{k_x}\epsilon_x, \notag\\
 l_1 \left( \partial_{k_y}\epsilon_x \right) &=& -\partial_{k_y}\epsilon_x, \notag \\
 l_1 \left( \partial_{k_y}\epsilon_y \right) &=& -\partial_{k_y}\epsilon_x.
\end{eqnarray}
Thus, $l_1v_x=v_x$, $l_1v_y=-v_y$, and then
\begin{eqnarray}
 l_1\hat{J}_x&=&e\sum_{\boldsymbol{k}}
(c_{\boldsymbol{k}_A}^\dagger~c_{\boldsymbol{k}_B}^\dagger~c_{-\boldsymbol{k}_A}~c_{-\boldsymbol{k}_B})
\begin{pmatrix}
0 & v_x & 0 & 0 \\
v_x^* & 0 & 0 & 0 \\
0 & 0 & 0 & v_x \\
0 & 0 & v_x^* & 0
\end{pmatrix}
\begin{pmatrix}
c_{\boldsymbol{k}_A} \\
c_{\boldsymbol{k}_B} \\
c_{-\boldsymbol{k}_A}^\dagger \\
c_{-\boldsymbol{k}_B}^\dagger
\end{pmatrix} \notag\\
&=& \hat{J}_x, \notag\\
 l_1\hat{J}_y&=&e\sum_{\boldsymbol{k}}
(c_{\boldsymbol{k}_A}^\dagger~c_{\boldsymbol{k}_B}^\dagger~c_{-\boldsymbol{k}_A}~c_{-\boldsymbol{k}_B})
\begin{pmatrix}
0 & -v_y & 0 & 0 \\
-v_y^* & 0 & 0 & 0 \\
0 & 0 & 0 & -v_y \\
0 & 0 & -v_y^* & 0
\end{pmatrix}
\begin{pmatrix}
c_{\boldsymbol{k}_A} \\
c_{\boldsymbol{k}_B} \\
c_{-\boldsymbol{k}_A}^\dagger \\
c_{-\boldsymbol{k}_B}^\dagger
\end{pmatrix} \notag\\
&=& -\hat{J}_y.
\end{eqnarray}

Under mirror reflection transformation along $y$-axis, i.e, the $l_4$-reflection defined in Fig. \ref{fig:C6vgroup} in the main text,
the above partial derivatives transform as
\begin{eqnarray}
 l_4 \left( \partial_{k_x}\epsilon_x \right) &=& -\partial_{k_x}\epsilon_x, \notag\\
  l_4 \left( \partial_{k_x}\epsilon_y \right) &=& \partial_{k_x}\epsilon_x, \notag\\
 l_4 \left( \partial_{k_y}\epsilon_x \right) &=& \partial_{k_y}\epsilon_x, \notag\\
  l_4 \left( \partial_{k_y}\epsilon_y \right) &=& -\partial_{k_y}\epsilon_x.
\end{eqnarray}
Thus, $l_4v_x=-v_x^*$, $l_4v_y=v_y^*$, and
\begin{eqnarray}
l_4\hat{J}_x&=&e\sum_{\boldsymbol{k}}
(c_{\boldsymbol{k}_B}^\dagger~c_{\boldsymbol{k}_A}^\dagger~c_{-\boldsymbol{k}_B}~c_{-\boldsymbol{k}_A})
\begin{pmatrix}
0 & -v_x^* & 0 & 0 \\
-v_x & 0 & 0 & 0 \\
0 & 0 & 0 & -v_x^* \\
0 & 0 & -v_x & 0
\end{pmatrix}
\begin{pmatrix}
c_{\boldsymbol{k}_B} \\
c_{\boldsymbol{k}_A} \\
c_{-\boldsymbol{k}_B}^\dagger \\
c_{-\boldsymbol{k}_A}^\dagger
\end{pmatrix} \notag\\
&=& -\hat{J}_x, \notag\\
 l_4\hat{J}_y&=&e\sum_{\boldsymbol{k}}
(c_{\boldsymbol{k}_B}^\dagger~c_{\boldsymbol{k}_A}^\dagger~c_{-\boldsymbol{k}_B}~c_{-\boldsymbol{k}_A})
\begin{pmatrix}
0 & v_y^* & 0 & 0 \\
v_y & 0 & 0 & 0 \\
0 & 0 & 0 & v_y^* \\
0 & 0 & v_y & 0
\end{pmatrix}
\begin{pmatrix}
c_{\boldsymbol{k}_B} \\
c_{\boldsymbol{k}_A} \\
c_{-\boldsymbol{k}_B}^\dagger \\
c_{-\boldsymbol{k}_A}^\dagger
\end{pmatrix} \notag\\
&=& \hat{J}_y.
\end{eqnarray}
Therefore, $l_1\pi_{xy}(\boldsymbol{q},\omega)=-\pi_{xy}(\boldsymbol{q},\omega)$,
$l_4\pi_{xy}(\boldsymbol{q},\omega)=-\pi_{xy}(\boldsymbol{q},\omega)$,
meaning that under mirror reflection transformation along $\hat{x}$- or $\hat{y}$-axis, the current correlator changes as $\pi_{xy} \to -\pi_{xy}$,  
which results in a sign change of $\sigma_H$.

However, if the mirror symmetry axis is not aligned with the $\hat{x}$- or $\hat{y}$-axis, but instead coincides with one of the other four axes defined in Fig. \ref{fig:C6vgroup} of the main text (e.g., the $l_2$-reflection),
the transformation of the aforementioned partial derivatives becomes
\begin{align}
& l_2 \left( \partial_{k_x}\epsilon_x \right) = ta\left[\sin \left( \frac{1}{2}k_xa+\frac{\sqrt{3}}{2}k_ya \right)-\frac{1}{2}\sin \left( \frac{1}{2}k_xa-\frac{\sqrt{3}}{2}k_ya \right)+\frac{1}{2}\sin \left( k_xa \right) \right], \notag\\
& l_2 \left( \partial_{k_x}\epsilon_y \right) = ta\left[\cos \left( \frac{1}{2}k_xa+\frac{\sqrt{3}}{2}k_ya \right)-\frac{1}{2}\cos \left( \frac{1}{2}k_xa-\frac{\sqrt{3}}{2}k_ya \right)-\frac{1}{2}\cos \left( k_xa \right) \right], \notag\\
& l_2 \left( \partial_{k_y}\epsilon_x \right) = -\frac{\sqrt{3}}{2}ta\left[-\sin \left( \frac{1}{2}k_xa-\frac{\sqrt{3}}{2}k_ya \right) - \sin \left( k_xa \right) \right], \notag\\
& l_2 \left( \partial_{k_y}\epsilon_y \right) = \frac{\sqrt{3}}{2}ta\left[\cos \left( \frac{1}{2}k_xa-\frac{\sqrt{3}}{2}k_ya \right) - \cos \left( k_xa \right) \right].
\end{align}
And the transformations under $l_3$-, $l_5$-, and $l_6$-reflections follow analogously.
Consequently, within the $C_{6v}$ point group, the four mirror reflection operations other than those along the $\hat{x}$- and $\hat{y}$-axes do not enforce a sign reversal of $\sigma_H$.
Therefore, a non-zero anomalous AC Hall conductivity can be realized without breaking the $l_2$-, $l_3$-, $l_5$-, and $l_6$-reflection symmetries.

\section{Anomalous AC Hall conductivity formulas}
\label{app:derive_Hall}

Perform an $S$-matrix expansion of the current-current correlator, Eq. \eqref{eq:pixy}, to one-loop level, we have
\begin{equation}
\pi_{xy}(\boldsymbol{q},\nu_m) = \frac{ie^2}{\beta} \sum_{\boldsymbol{k},\omega_n} \text{Tr} 
\left[ \hat{v}_x \left(\boldsymbol{k}+\frac{\boldsymbol{q}}{2} \right) G_0(\boldsymbol{k},\omega_n) \hat{v}_y \left(\boldsymbol{k}+\frac{\boldsymbol{q}}{2} \right) G_0(\boldsymbol{k}+\boldsymbol{q},\omega_n+\nu_m) \right],
\label{eq:pixyqv}
\end{equation}
where $G_0(\boldsymbol{k},\omega_n)=\left( i\hbar \omega_n - h_{\boldsymbol{k}} \right)^{-1}$,
and $\omega_n$ is the fermionic Matsubara frequency, satisfying $\hbar \omega_n=(2n+1)\pi/\beta$.
Take the antisymmetric difference of \eqref{eq:pixyqv} and simplify for the $\boldsymbol{q}=0$ case of interest,
\begin{align}
\pi_{xy}(\nu_m)-\pi_{yx}(\nu_m) = \frac{ie^2}{\beta} \sum_{\boldsymbol{k},\omega_n}
\frac{4\hbar \nu_m(\hbar \nu_m+2\hbar \omega_n)^2\left( v_x^*v_y-v_xv_y^* \right) \left[ \mu \left( \Delta_x^*\Delta_y-\Delta_x\Delta_y^* \right) + 2i\Delta_s \left( \epsilon_x \text{Im}\Delta_y-\epsilon_y \text{Im}\Delta_x \right) \right]}{\left( \hbar^2 \omega_n^2+E_1^2 \right) \left( \hbar^2 \omega_n^2+E_2^2 \right) \left[ \hbar^2 (\omega_n+\nu_m)^2+E_1^2 \right] \left[ \hbar^2 (\omega_n+\nu_m)^2+E_2^2 \right]},
\end{align}
Thus, the anomalous AC Hall conductivity can be written as
\begin{align}
\sigma_H(\nu_m) &= \frac{1}{2\hbar\omega} \left[ \pi_{xy}(\nu_m)-\pi_{yx}(\nu_m) \right] \notag\\
&= \frac{e^2}{\hbar\beta} \int \frac{d^2k}{(2\pi)^2} \sum_{\omega_n}
\frac{2\hbar(\hbar \nu_m+2\hbar \omega_n)^2\left( v_x^*v_y-v_xv_y^* \right) \left[ \mu \left( \Delta_x^* \Delta_y - \Delta_x \Delta_y^* \right) + 2i\Delta_s \left( \epsilon_x \text{Im} \Delta_y -\epsilon_y\text{Im} \Delta_x \right) \right]}{\left( \hbar^2 \omega_n^2+E_1^2 \right) \left( \hbar^2 \omega_n^2+E_2^2 \right) \left[ \hbar^2 (\omega_n+\nu_m)^2+E_1^2 \right] \left[ \hbar^2 (\omega_n+\nu_m)^2+E_2^2 \right]}.
\end{align}
Taking the limit $i\nu_m \to \omega+i\epsilon$, the above equation corresponds to Eq. \eqref{eq:sigmaH} in the main text.
By performing the Matsubara summation over $\omega_n$, we obtain
\begin{align}
\sigma_H(\omega) = & \frac{e^2}{\hbar} \int \frac{d^2k}{\left(2\pi \right)^2} \hbar^2\left( v_x^*v_y-v_xv_y^* \right) \left[ \mu \left( \Delta_x^* \Delta_y - \Delta_x \Delta_y^* \right) + 2i\Delta_s \left( \epsilon_x \text{Im} \Delta_y -\epsilon_y\text{Im} \Delta_x \right) \right] \times \notag\\
& \left \{ \frac{1-n_F(E_1)-n_F(E_2)}{2E_1E_2(E_1+E_2)^2} \left[ \frac{E_1+E_2+\hbar \omega -i\epsilon}{(E_1+E_2+\hbar \omega)^2+\epsilon^2} + \frac{E_1+E_2-\hbar \omega +i\epsilon}{(E_1+E_2-\hbar \omega)^2+\epsilon^2} \right] \right. \notag\\
&- \left. \frac{n_F(E_1)-n_F(E_2)}{2E_1E_2(E_1-E_2)^2} \left[ \frac{E_1-E_2+\hbar \omega -i\epsilon}{(E_1-E_2+\hbar \omega)^2+\epsilon^2} + \frac{E_1-E_2-\hbar \omega +i\epsilon}{(E_1-E_2-\hbar \omega)^2+\epsilon^2} \right] \right \},
\label{eq:sigmaHw}
\end{align}
where $\epsilon$ is a positive infinitesimal.
At $T=0K$, $n_F(E_1)=n_F(E_2)=0$, and $\left[ \sigma_H(\omega) \right]_{T=0K}$ can be written as
\begin{align}
\left[\sigma_H(\omega) \right]_{T=0K} = & \frac{e^2}{\hbar} \int \frac{d^2k}{\left(2\pi \right)^2} \hbar^2 \left( v_x^*v_y-v_xv_y^* \right) \left[ \mu \left( \Delta_x^* \Delta_y - \Delta_x \Delta_y^* \right) + 2i\Delta_s \left( \epsilon_x \text{Im} \Delta_y -\epsilon_y\text{Im} \Delta_x \right) \right] \times \notag\\
& \frac{1}{2E_1E_2(E_1+E_2)^2} \left[ \frac{E_1+E_2+\hbar \omega -i\epsilon}{(E_1+E_2+\hbar \omega)^2+\epsilon^2} + \frac{E_1+E_2-\hbar \omega +i\epsilon}{(E_1+E_2-\hbar \omega)^2+\epsilon^2} \right].
\label{eq:sigmaH0K}
\end{align}

\section{Supercurrent of the tri-component pairing superconductor in magnetic fields}
\label{app:supercurrent}

It is not straightforward to infer the vortex configurations directly from Eq. \eqref{eq:F_A} in the main text.
As pointed out in Ref. \onlinecite{Babaev2002}, the GL free energy functional of a two-gap superconductor can be exactly mapped onto the extended Faddeev model, which consists of a three-component unit vector $\boldsymbol{n}$, a massive vector field $\mathbf{C}$, and a density-related variable $\rho$.
In the tri-component superconducting system with $d_{x^2-y^2}$-, $d_{xy}$-, and $s$-wave pairings,
$\boldsymbol{n}$ is extended to an eight-component vector in the $SU(3)$ case,
defined in terms of the normalized complex order-parameter vector $\boldsymbol{z}$ and the $SU(3)$ Gell-Mann matrices $\lambda_a (a=1,2,...,8)$:
\begin{equation}
n_a=\boldsymbol{z}^\dagger \lambda_a \boldsymbol{z},
\label{eq:n_a}
\end{equation}
in which
\begin{equation}
\boldsymbol{z}=\frac{1}{\rho}\left( \frac{\psi_s}{\sqrt{2m_s}}~\frac{\psi_1}{\sqrt{2m_1}}~\frac{\psi_2}{\sqrt{2m_2}} \right)^T,
\label{eq:z}
\end{equation}
and the order parameters can be written as
\begin{align}
& \psi_s=\sqrt{2m_s}\rho \cos{\left(\frac{\theta}{2}\right)} e^{i\phi_s}, \notag\\
& \psi_1=\sqrt{2m_1}\rho \sin{\left(\frac{\theta}{2}\right)} \cos{\left(\frac{\gamma}{2}\right)} e^{i\phi_1}, \notag\\
& \psi_2=\sqrt{2m_2}\rho \sin{\left(\frac{\theta}{2}\right)} \sin{\left(\frac{\gamma}{2}\right)} e^{i\phi_2}.
\label{eq:psi_rho}
\end{align}

Substituting Eq. \eqref{eq:psi_rho} into Eq. \eqref{eq:F_A} and adopting the London limit ($\left|\psi_{s,1,2}\right|=$const), we obtain
\begin{equation}
F = \frac{\rho^2}{4} \sum_{a} \left( \nabla n_a \right)^2+\frac{\rho^2}{16}\mathbf{C}^2+\frac{\mathbf{B}^2}{2}+V_0+V_{\rho Kn},
\label{eq:F_Faddeev}
\end{equation}
where
\begin{equation}
\mathbf{C}=\frac{i}{m_s\rho^2}\left(\psi_s^* \nabla \psi_s-\psi_s \nabla \psi_s^* \right)+\frac{i}{m_1\rho^2}\left(\psi_1^* \nabla \psi_1-\psi_1 \nabla \psi_1^* \right)+\frac{i}{m_2\rho^2}\left(\psi_2^* \nabla \psi_2-\psi_2 \nabla \psi_2^* \right)-\frac{4e}{\rho^2}\left(\frac{\left|\psi_s\right|^2}{m_s}+\frac{\left|\psi_1\right|^2}{m_1}+\frac{\left|\psi_2\right|^2}{m_2} \right)\mathbf{A},
\label{eq:C}
\end{equation}
\begin{equation}
V_0=F_s^{(0)}+F_d^{(0)}-2g_{dd}|\psi_1|^2|\psi_2|^2,
\label{eq:V_0}
\end{equation}
and
\begin{equation}
V_{\rho Kn}=\rho^4 \left[K_1\left(n_1^2-n_2^2 \right)+K_2\left(n_4^2-n_5^2 \right)+K_3\left(n_6^2-n_7^2 \right)+K_4n_1n_3^\prime+K_5n_4n_6 \right],
\label{eq:V_rhoKn}
\end{equation}
in which $K_1=2g_{sd}m_sm_1$, $K_2=2g_{sd}m_sm_2$, $K_3=2g_{dd}m_1m_2$, $K_4=4g_{sd}^\prime \sqrt{m_sm_1}$,
$K_5=4g_{sd}^\prime m_2\sqrt{m_sm_1}$, and
\begin{align}
& n_1=\sin{\theta}\cos{\left(\frac{\gamma}{2} \right)} \cos{\delta_1}, ~ n_2=\sin{\theta}\cos{\left(\frac{\gamma}{2} \right)}\sin{\delta_1}, \notag\\
& n_3=\sin^2{\left(\frac{\theta}{2}\right)}\cos{\gamma}, ~ n_3^\prime=\sin^2{\left(\frac{\theta}{2}\right)}\left[m_1\cos^2{\left(\frac{\gamma}{2} \right)}-m_2\sin^2{\left(\frac{\gamma}{2} \right)} \right], \notag\\
& n_4=\sin{\theta}\sin{\left(\frac{\gamma}{2} \right)} \cos{\delta_2}, ~ n_5=\sin{\theta}\sin{\left(\frac{\gamma}{2} \right)} \sin{\delta_2}, \notag\\
& n_6=\sin^2{\left(\frac{\theta}{2}\right)}\sin{\gamma} \cos{\delta_0}, ~ n_7=\sin^2{\left(\frac{\theta}{2}\right)}\sin{\gamma} \sin{\delta_0}, \notag\\
& n_8=\frac{1}{\sqrt{3}}\left[\cos^2{\left(\frac{\theta}{2}\right)}+\sin^2{\left(\frac{\theta}{2}\right)}\cos^2{\left(\frac{\gamma}{2}\right)}-2\sin^2{\left(\frac{\theta}{2}\right)}\cos^2{\left(\frac{\gamma}{2}\right)} \right],
\label{n_a_1_8}
\end{align}
where $\gamma_{0,1,2}$ denote the relative phase differences among the three order parameters, with
$\gamma_0=\phi_1-\phi_2$, $\gamma_1=\phi_1-\phi_s$, and $\gamma_2=\phi_2-\phi_s$.
The free energy functional can be further expressed as:
\begin{align}
F=& \frac{\rho^2}{4}\left[ \sin^2{\theta} \cos^2\left(\frac{\gamma}{2}\right) \left(\nabla \gamma_1 \right)^2
  + \sin^2{\theta} \sin^2\left(\frac{\gamma}{2}\right) \left(\nabla \gamma_2 \right)^2
  + \sin^4\left(\frac{\theta}{2}\right) \sin^2{\gamma} \left(\nabla \gamma_0 \right)^2 \right] \notag\\
 &+ \rho^2 \left[ \sin^2{\left(\frac{\theta}{2}\right)} \cos^2{\left(\frac{\gamma}{2}\right)} \nabla \phi_1
   +\sin^2{\left(\frac{\theta}{2}\right)} \sin^2{\left(\frac{\gamma}{2}\right)} \nabla \phi_2
   +\cos^2{\left(\frac{\theta}{2}\right)} \nabla \phi_s + 2e\mathbf{A} \right]^2 + \frac{\mathbf{B}^2}{2} + V_0 +V_{\rho Kn}.
\label{eq:F_rho}
\end{align}
The variation of the second term in the free energy functional given in Eq. \eqref{eq:F_rho} directly leads to the expression for the supercurrent:
\begin{equation}
\mathbf{J}=-4e\rho^2 \left[ \cos^2{\left(\frac{\theta}{2}\right)} \nabla \phi_s+\sin^2{\left(\frac{\theta}{2}\right)} \cos^2{\left(\frac{\gamma}{2}\right)} \nabla \phi_1 + \sin^2{\left(\frac{\theta}{2}\right)} \sin^2{\left(\frac{\gamma}{2}\right)} \nabla \phi_2+2e\mathbf{A} \right].
\end{equation}

\end{widetext}

\end{document}